\documentclass[12pt,a4]{article}

\newif\ifpdf 
\ifx\pdfoutput\undefined 
\else 
\pdfoutput=1 
\pdftrue 
\fi

\ifpdf
\usepackage[pdftex]{color}
\usepackage[pdftex]{graphicx}
\else
\usepackage[dvips]{color}
\usepackage[dvips]{graphicx}
\fi

\newcommand{\fig}{.}
\newcommand{\tverFig}{.}
\newcommand{\slideFig}{.}
\newcommand{\tdrFig}{.}
\newcommand{\un}[2]{\ensuremath{\mathrm{#1 \, #2}}}
\newcommand{\units}[1]{\ensuremath{\mathrm{#1}}}
\newcommand{\degrees}[1]{\ensuremath{\mathrm{#1^{\circ}}}}
\newcommand{\chem}[1]{\ensuremath{\mathrm{#1}}}

\newcommand{\prt}[1]{\ensuremath{{\rm #1}}}
\newcommand{\qrk}[1]{\ensuremath{#1}}
\newcommand{\Bo}{\prt{B^0}}

\newcommand{\Bso}{\prt{B_{s}^0}}

\newcommand{\Bdo}{\prt{B_{d}^0}}
\newcommand{\Bdob}{\prt{\overline{B_{d}^0}}}

\newcommand{\Jpsi}{\prt{J\!/\!\psi}}
\newcommand{\Ks}{\prt{K_S}}
\newcommand{\pip}{\prt{\pi^+}}
\newcommand{\pim}{\prt{\pi^-}}

\newcommand{\Bpipi}{\prt{\Bdo \to \pip \pim}}

\newcommand{\BJpsiK}{\prt{\Bdo \to \Jpsi \Ks}}

\newcommand{\gam}{\ensuremath{\gamma}}
\newcommand{\bet}{\ensuremath{\beta}}

\newcommand{\bbar}{\qrk{b\overline{b}}}
\newcommand{\dm}{\ensuremath{\Delta\!m}}

\newcommand{\C}{Cherenkov}

\newcommand{\tick}{\ensuremath{\mathbf{\surd}}}
\newcommand{\np}{New Physics}
\newcommand{\sm}{Standard Model}

\newcommand{\cp}{\ensuremath{\mathcal{CP}}}

\newcommand{\lhcb}{{L}{H}{C}{b}}
\newcommand{\lhc}{{L}{H}{C}}

\newcommand{\pythia}{PYTHIA}
\newcommand{\pt}{\ensuremath{\mathrm{p_{t}}}}

\newcommand{\Adirpp}{\ensuremath{A_{\prt{\pi\pi}}^{\mathrm{Dir}}}}
\newcommand{\Amixpp}{\ensuremath{A_{\prt{\pi\pi}}^{\mathrm{Mix}}}}
\newcommand{\AdirKK}{\ensuremath{A_{\prt{KK}}^{\mathrm{Dir}}}}
\newcommand{\AmixKK}{\ensuremath{A_{\prt{KK}}^{\mathrm{Mix}}}}
\newcommand{\phis}{\ensuremath{2\delta\gamma}}

\newcommand{\dms}{\ensuremath{\Delta m_s}}
\newcommand{\invps}{\units{ps^{-1}}}
\newcommand{\DGsGs}{\ensuremath{\frac{\Delta \Gamma_s}{\Gamma_s}}}
\newcommand{\djdj}{\ensuremath{2\delta\gamma + \gamma}}
\newcommand{\UT}{Unitarity Triangle}
\newcommand{\lzero}{Level-0}
\newcommand{\lone}{Level-1}
\newcommand{\hlt}{HLT}

\newcommand{\mod}[1]{\ensuremath{\left| #1 \right|}}
\newcommand{\mIII}[9]{\ensuremath{%
  \left( \begin{array}{ccc} #1 & #2 & #3 \\ 
                           #4 & #5 & #6 \\
                           #7 & #8 & #9 \end{array} \right)}}

\newcommand{\order}[1]{\ensuremath{\mathcal{O}\!\left(#1\right)}}

\newcommand{\nd}{\ensuremath{\mathrm{\mbox{}^{nd}}}}

\newcommand{\captionFont}{\sf}

\begin{document}
\begin{center}
{\bf {\Large LHCb: Status and Physics Prospects }\\
 {\small Contribution to the
 XVIIth International workshop on high energy physics
 and quantum field theory (QFTHEP'03) Samara-Saratov, 
 Russia, Sept 4-11, 2003.}
\vspace{4 mm}

}
            Jonas Rademacker on behalf of the LHCb Collaboration

\vspace{4 mm}

University of Oxford \\ 
Denys Wilkinson Bldg, Keble Road, Oxford OX1~3RH, UK\\

\end{center}

\begin{abstract}
 We discuss the current status and the physics prospects at the LHCb
 detector, the dedicated B physics detector at the LHC, due to start
 data taking in 2007.
\end{abstract}

\section{Introduction}                      
 LHCb is a dedicated B--physics experiment at the future LHC collider,
 making use of the large number of B--hadrons expected at the LHC. The
 experiment is scheduled to start data taking in 2007. Here we will
 introduce the LHCb detector, and its physics potential, focusing on
 one of its most exciting features, LHCb's ability to perform
 precision measurements on the CKM angle \gam\ in many different decay
 channels, in both the \Bso\ and the \Bdo\ system. This will
 thoroughly over constrain the \sm\ description of CP violation and
 provide a sensitive probe for \np.

 A more detailed description of the LHCb detector and its projected
 physics performance can be found in the LHCb technical design reports
 \cite{tdr}.
\section{CP Violation}
\subsection{CP Violation in the \sm}
 In the \sm, \cp\ violation can be accommodated by a single complex
 phase $\delta_{13}$ in the CKM matrix, which is the matrix that
 relates the mass--eigenstates of the down--type quarks to the weak
 isospin partners of the up--type quarks:
\begin{equation}
\label{eq:th.b.ckmlabel}
V_{\mathrm{CKM}}=\mIII{ V_{ud} }{ V_{us} }{ V_{ub} }{
                        V_{cd} }{ V_{cs} }{ V_{cb} }{
                        V_{td} }{ V_{ts} }{ V_{tb} }
.
\end{equation}
 The transition amplitudes between quarks are proportional to the
 corresponding elements in the CKM matrix, for example the amplitude
 for \qrk{d_L \to u_L} is proportional to $V_{ud}$, while the
 CP-conjugate process, \qrk{\bar{d}_R \to \bar{u}_R} is proportional
 to the complex conjugate, $V_{ud}^{\ast}$.  Experimentally, it is
 found that the magnitudes of the CKM matrix elements follow a clear
 structure. In terms of the sine of the Cabibbo angle, $\lambda\equiv
 \sin\theta_C=0.22$, the order of magnitude of the CKM matrix elements
 is:
\begin{equation}
\mIII{ 1 }{\lambda}{\lambda^3}{
      \lambda}{1}{\lambda^2}{
      \lambda^3}{\lambda^2}{1}
.
\end{equation}\\
 Up to \order{\lambda^3} in the Wolfenstein parametrisation of the CKM
 matrix \cite{Wolfenstein:1983yz}, only the two smallest elements have
 complex phases (these phases are not independent and would vanish if
 $\delta_{13}$ were 0):
\begin{equation}
 V_{td} = \mod{V_{td}}e^{-i\bet} \mbox{\hspace{3em} and \hspace{3em}}
 V_{ub} = \mod{V_{ub}}e^{-i\gam} 
.
\label{eq:introBetaGamma}
\end{equation}
 At \order{\lambda^4}, another, phase appears, $\delta\gamma$:
\begin{equation}
 V_{ts} = \mod{V_{ts}}e^{-i\delta\gamma}.
\end{equation}
 All three phases up to \order{\lambda^4}, \bet, \gam\ and
 $\delta\gamma$, are accessible in B-systems. The phase \gam\ appears
 in all decays involving \qrk{b \to u} transitions, for example
 \prt{B_d \to \pi^+\pi^-} and \prt{B_s \to K^+K^-}. The phase \bet\
 appears in \prt{B_d} mixing, where a \prt{B_d} meson transforms into
 a \prt{\bar{B}_d} meson: \( \prt{B_d}
 \stackrel{-2\beta}{\longrightarrow} \prt{\bar{B}_d} \). Analogously,
 the phase $\delta\gamma$ is the mixing angle of the \prt{B_s} system,
 \(
 \prt{B_s} \stackrel{-2\delta\gamma}{\longrightarrow} \prt{\bar{B}_s}
 \).  
 While \bet\ and \gam\ are \order{1}, $\delta\gamma$ is expected
 to be \order{10^{-2}} in the Standard Model.

\begin{figure}
\begin{center}
\begin{tabular}{cc}
(a) \BJpsiK & (b) \Bpipi \\
\parbox{0.4\textwidth}{
  \center\includegraphics[width=0.3\textwidth]{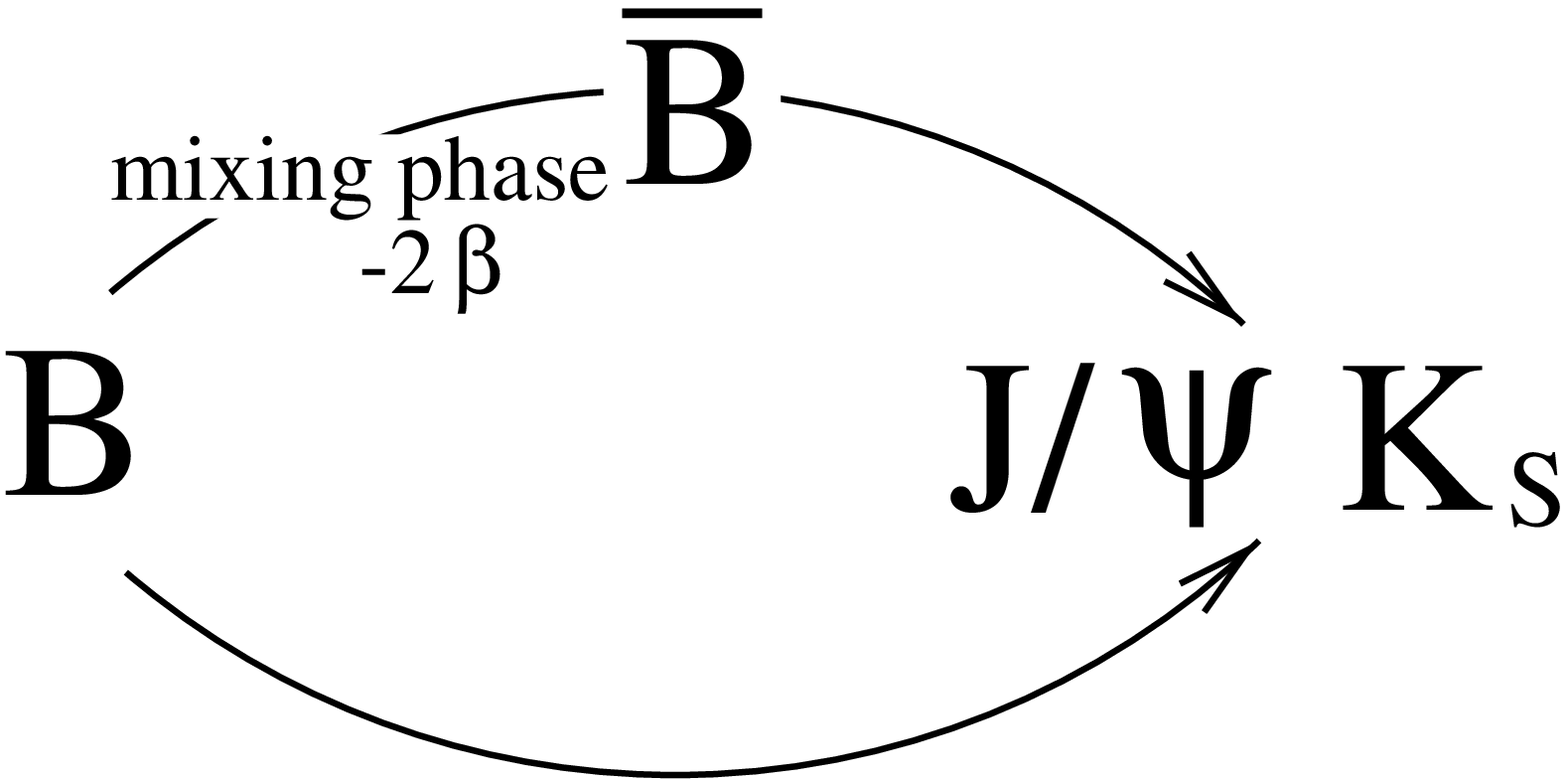}
}
&
\parbox{0.4\textwidth}{
  \center\includegraphics[width=0.3\textwidth]{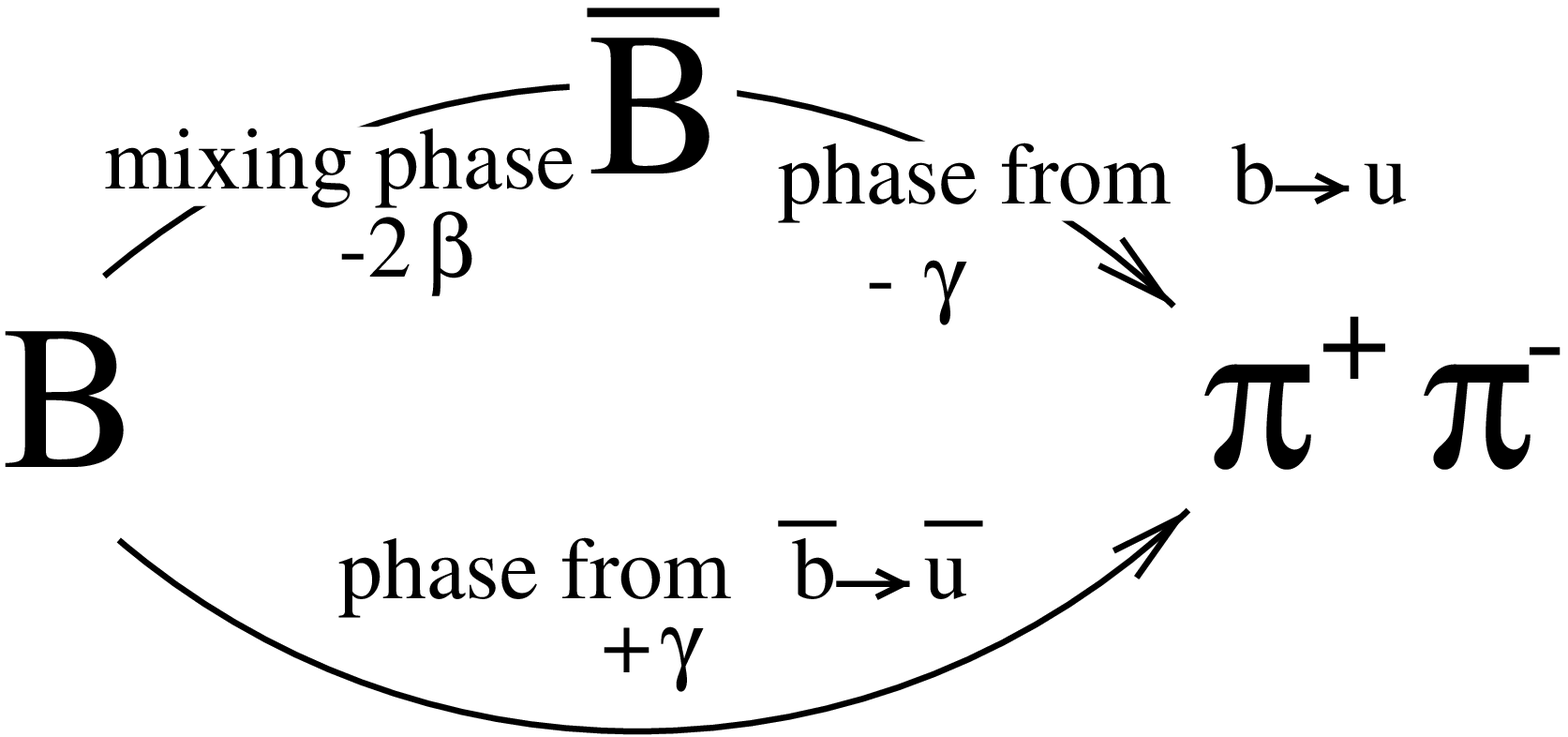}
}
\\
$\delta_{\mathrm{CP}}= -2\bet$ & $\delta_{\mathrm{CP}}=
-2\left(\bet+\gam\right)$
\end{tabular}
\caption{\captionFont The CKM phases are observable as phase differences between
 interfering decay paths to the same final state. Here illustrated for
 the examples \BJpsiK\ (sensitive to $\sin2\beta$) and \Bpipi\
 (sensitive to $\sin(2\beta + 2\gamma)$). This illustration ignores
 the penguin contributions to \Bpipi, which are discussed later in
 the text.
 \label{interference}}
\end{center}
\end{figure}

 The complex CKM elements result in phase differences between
 interfering decay paths to the same final state, one with and one
 without mixing, as illustrated in figure \ref{interference}.
 These phase differences can be observed as the amplitudes of time
 dependent decay rate asymmetries, for example for \prt{B
 \to J\!/\!\psi K_s}:
\begin{equation}
\label{eq:th.b.asybeta}
 A\!\left(\tau\right)=\frac{\Gamma\left(\prt{\Bdo\to \Jpsi K_S}\right)
-\Gamma\left(\prt{\Bdob\to \Jpsi K_S}\right)}{
\Gamma\left(\prt{\Bdo\to \Jpsi K_S}\right)
+\Gamma\left(\prt{\Bdob\to \Jpsi K_S}\right)}
 = \sin\!\left(2\bet \right) \sin\!\left(\dm \tau \, \right)
,
\end{equation}\\
 where \dm\ is the mass difference between the two \Bdo\ mass
 eigenstates, $\tau$ is the decay eigentime and the flavours \Bdo\ and
 \Bdob\ refer to the flavour at the time of creation ($\tau = 0$).  An
 experiment measuring CP violation in the B systems would therefore
 require a good time (decay length) resolution, especially to resolve
 the rapid \prt{B_s} oscillations. Also, because the branching
 fractions to CP sensitive decays are typically \order{10^{-5}}, large,
 clean data samples are required, and efficient B flavour tagging (the
 ability to identify the flavour of the B at the time of
 creation). LHCb is specifically designed to meet these requirements.

\subsection{\UT}
\label{sec:th.b.status2006}
 The only Standard Model prediction with respect to the CKM matrix is
 that it is unitary:
\begin{equation}
 V_{\mathrm{CKM}} V_{\mathrm{CKM}}^{\dagger} = 1\!\!1
\end{equation}
 This results in 9 equations. The most relevant one for CP violation
 in the B systems is
\begin{equation}
 V_{ub}^{\ast}V_{ud}+V_{cb}^{\ast}V_{cd}+V_{tb}^{\ast}V_{td}=0.
\end{equation}
 which can also be written as:
\begin{equation}
   \frac{V_{ub}^{\ast}V_{ud}}{V_{cb}^{\ast}V_{cd}}
 + 1
 + \frac{V_{tb}^{\ast}V_{td}}{V_{cb}^{\ast}V_{cd}}
 = 0.
\label{eq:unitaryNormal}
\end{equation}
 Drawing these three numbers adding up to zero as points in the
 complex plane, results in the \UT.
\begin{figure}
\begin{center}
\begin{tabular}{cc}
\includegraphics[width=0.4\textwidth]{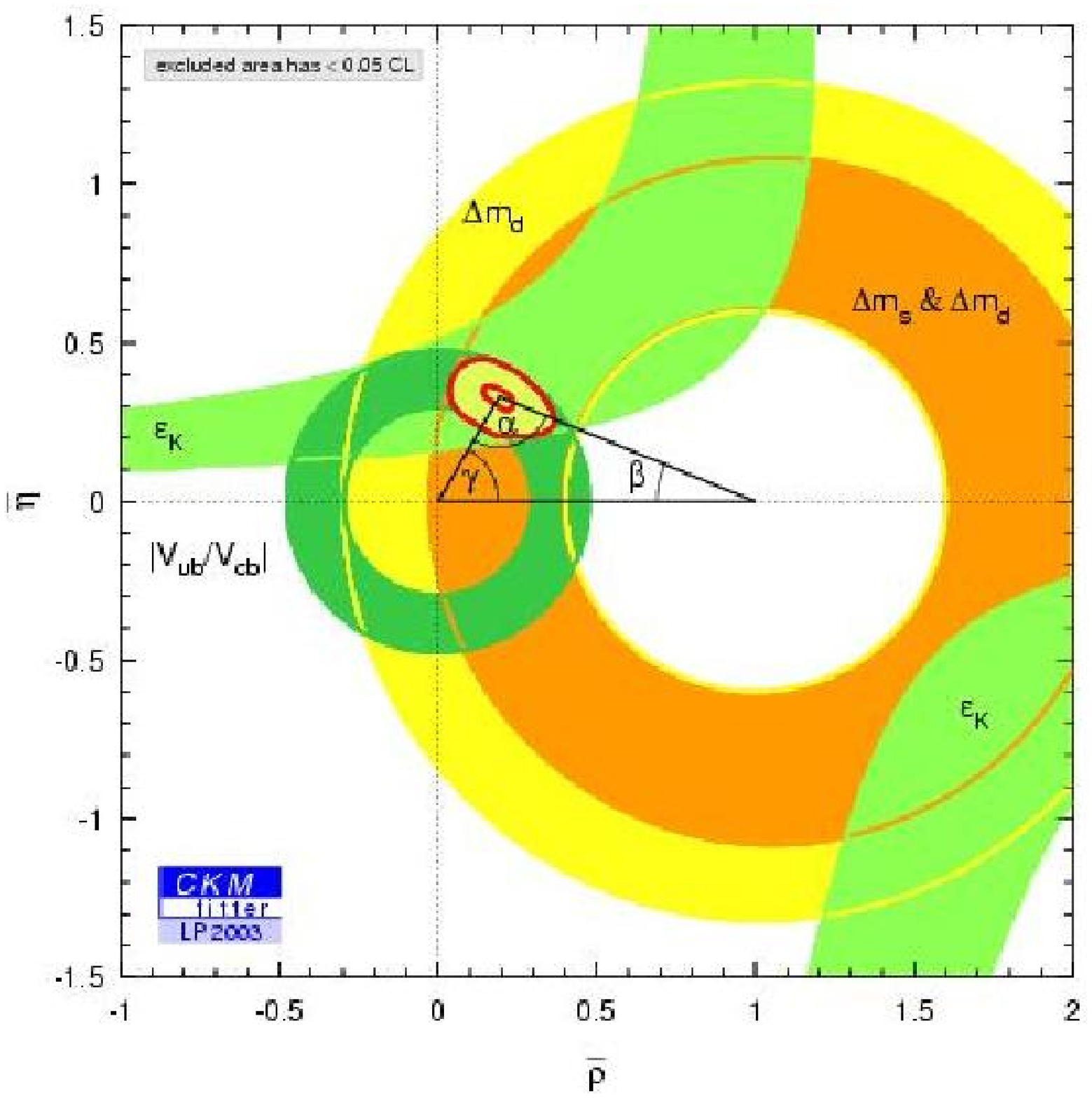}
&
\includegraphics[width=0.4\textwidth]{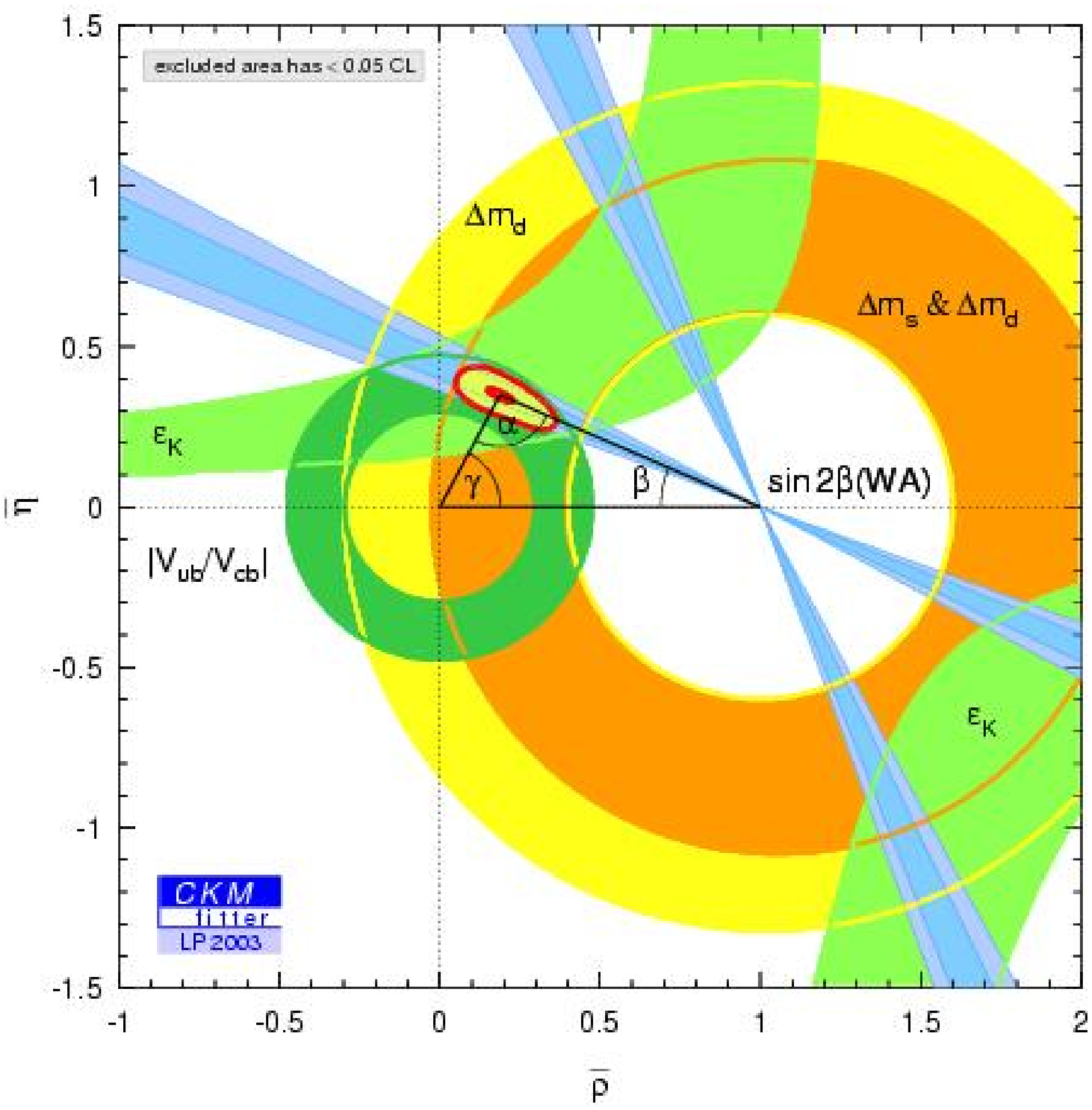}\\
 Without direct measurement &  Including $\sin 2 \beta$ measurement\\
  of $\sin 2 \beta$         &  from charmonium decays \cite{hfag}.
\end{tabular}
\caption{\captionFont Current status of the \UT~\cite{ckmfitter}. 
  The $\bar{\rho}-\bar{\eta}$ plane is effectively
  the complex plane, using parameters of the improved
  \cite{Buras:1994ec} Wolfenstein \cite{Wolfenstein:1983yz}
  parametrisation. The angles $\beta$ and
  $\gamma$ correspond to the CP-violating phases introduced in
  Eq \ref{eq:introBetaGamma}. The third angle $\alpha$ does not
  correspond to another complex entry in the CKM matrix, and is
  defined as $\alpha = \pi - \beta - \gamma$.\label{fig:triangles}}
\end{center}
\end{figure}
 The normalised \UT\ (Eq \ref{eq:unitaryNormal}) is fully described by
 the position of its apex in the complex plane. The angles correspond
 to the CKM-phases $\beta, \gamma$, introduced above. The third angle
 $\alpha$, often found in the literature, is given by $\alpha = \pi -
 \beta - \gamma$. The \UT\ provides an elegant way to relate the
 phases to other measurements that determine the sides of the \UT.
 The \UT, and current constraints on the position of its apex, is
 shown in figure \ref{fig:triangles}.  Combining direct measurement of
 $\sin 2\beta$ from oscillation experiments (dominated by \prt{B_d \to
 J/\psi K_s} at BaBar and BELLE), restricted to the charmonium
 results, the Heavy Flavour Averaging Group find \cite{hfag}.
\begin{equation}
 \sin 2\beta = 0.736 \pm 0.049
\end{equation}
 From a global fit to the \UT, ignoring the direct
 $\sin 2\beta$ measurements from B oscillations, the CKM-Fitter group
 find \cite{ckmfitter}
\begin{equation}
  \sin 2\beta = 0.587 - 0.766\;\mbox{at $68\%$ confidence level}
\end{equation}
 in excellent agreement. However, the angle $\gamma$ has not yet been
 measured directly.

 By the year $2007$, the accuracy of both the side measurements and
 direct measurements of $\sin 2 \beta$ will have increased
 significantly. While first estimates of \gam\ might be possible, the
 uncertainties are expected to be too large to give strong constraints
 on the \sm\ description of \cp\ violation.

\section{The LHCb experiment}

\subsection{Bottom Production at the LHC}
\label{sec:bottomsAtLHC}
 The planned Large Hadron Collider (LHC) at CERN will collide protons
 at a centre--of--mass energy of \un{14}{TeV} at a design luminosity
 of \un{\sim 10^{34}}{cm^{-1}s^{-1}} at the high-luminosity
 interaction points (ATLAS and CMS).
 The accelerator will be housed in the \un{27}{km} tunnel that has
 been built for the LEP experiment. \lhc\ is scheduled to start data
 taking in 2007 with a luminosity of \un{10^{33}}{cm^{-1}s^{-1}} and
 upgrade to its full luminosity after a few years.
 Due to the huge \qrk{b} production cross section of \un{\sim 500}{\mu
 b} \cite{yellowbproduction}, \lhc\ will be the most copious source of
 \prt{B} hadrons in the world by several orders of magnitude.

\begin{figure}
\begin{center}
\begin{tabular}{ccc}
\parbox{0.2\textwidth}{
  \includegraphics[width=0.2\textwidth]{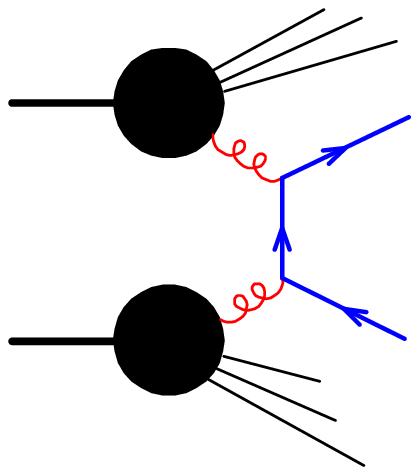}}
&
\parbox{0.3\textwidth}{\includegraphics[width=0.3\textwidth]{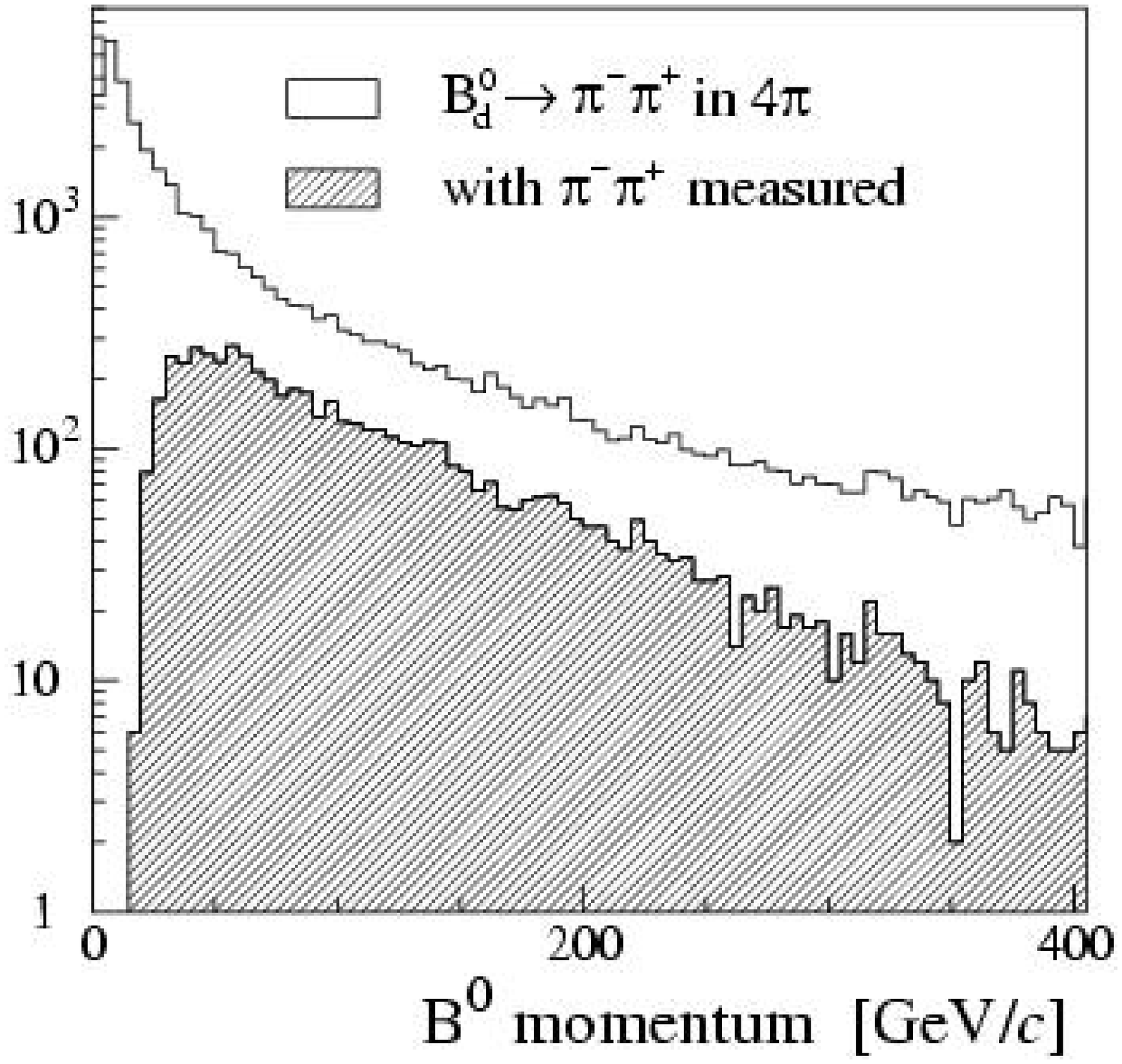}}
&
\parbox{0.3\textwidth}{\includegraphics[width=0.3\textwidth]{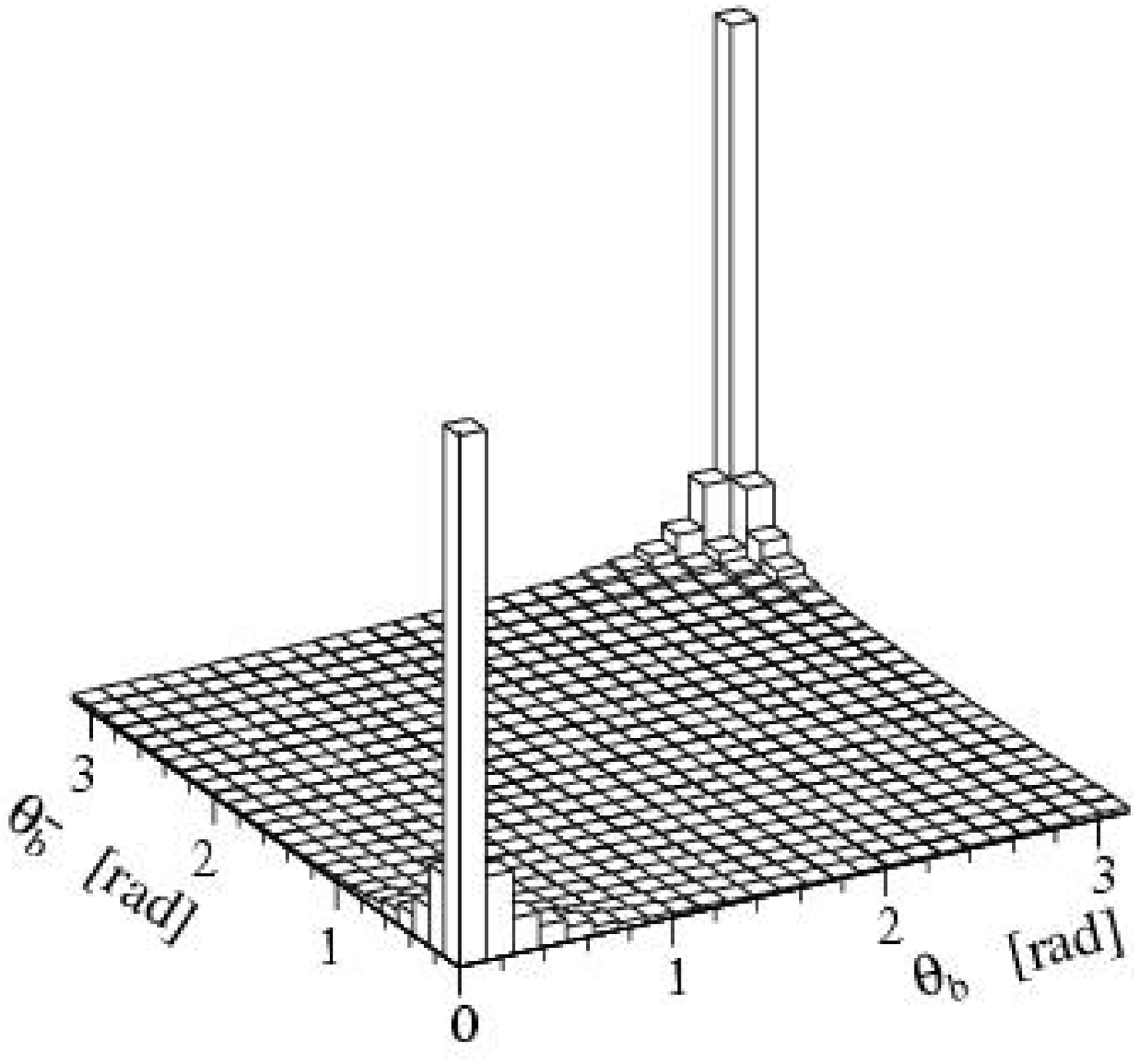}}
\\
\parbox{0.22\textwidth}{(a)Typical diagram for
 \bbar\ production in pp collisions
 \cite{yellowbproduction}.}
&
\parbox{0.35\textwidth}{ (b)
 Momentum distribution for \Bdo\ mesons in \prt{\Bdo\to\pi^+\pi^-}
 decays \cite{lhcb:tp}.} &
\parbox{0.35\textwidth}{ (c)
 Polar angles of hadrons formed from \bbar\ pairs, calculated by
 \pythia\ \cite{lhcb:tp}. }\\
\end{tabular}
\caption{\captionFont B--hadrons at \lhcb. \label{fig:det.lhcb.bmomenta}}
\end{center}
\end{figure}

 The kinematics of B hadron production at \un{14}{TeV} \prt{p-p}, as
 illustrated in figure \ref{fig:det.lhcb.bmomenta}, have major
 consequences of the design of a dedicated B physics detector:
\begin{itemize}
\item The B hadrons produced are highly boosted, which results in
 long decay lengths (\un{\sim 1}{cm}) and hence facilitates exact decay
 time measurements.
\item Both, the \qrk{b} and the \qrk{\bar{b}}, are predominantly
 produced in the same forward or backward cone, so that a single--arm
 spectrometer captures both B--hadrons produced, which is essential
 for \Bo--tagging, as discussed in Section \ref{sec:bTagging}.
\end{itemize}

\subsection{Luminosity at the LHCb interaction point}

 At the \lhc\ design luminosity, each bunch crossing would involve many
 inelastic proton--proton interaction. Such multiple interactions
 severely complicate the task of \Bo--tagging, and of cleanly locating
 the primary and secondary vertices.

 Therefore the luminosity at the \lhcb\ detector is reduced to
 \un{2\cdot 10^{32}}{cm^{-1}s^{-1}} by defocussing the beam at the
 \lhcb\ interaction point. Apart from optimising the number of single
 interactions, also the trigger performance, detector occupancy and
 radiation levels are taken into account when choosing the design
 luminosity.
 Remaining multiple interactions are identified by the Pile-Up
 system. While LHCb is optimised for single interactions, remaining
 multiple interactions are not necessarily discarded. The decision
 whether to keep of discard a multiple interaction event is made at
 trigger \lzero.

 With this luminosity, \lhcb\ expects about $10^{12}$ \bbar\ events
 per year. Due to its comparably moderate luminosity requirements,
 \lhcb\ can start its full physics programme from the first day of
 \lhc\ running.

\subsection{The LHCb detector}
\label{sec:det.lhcb}
\begin{figure}
\begin{center}
\includegraphics[width=0.99\textwidth]{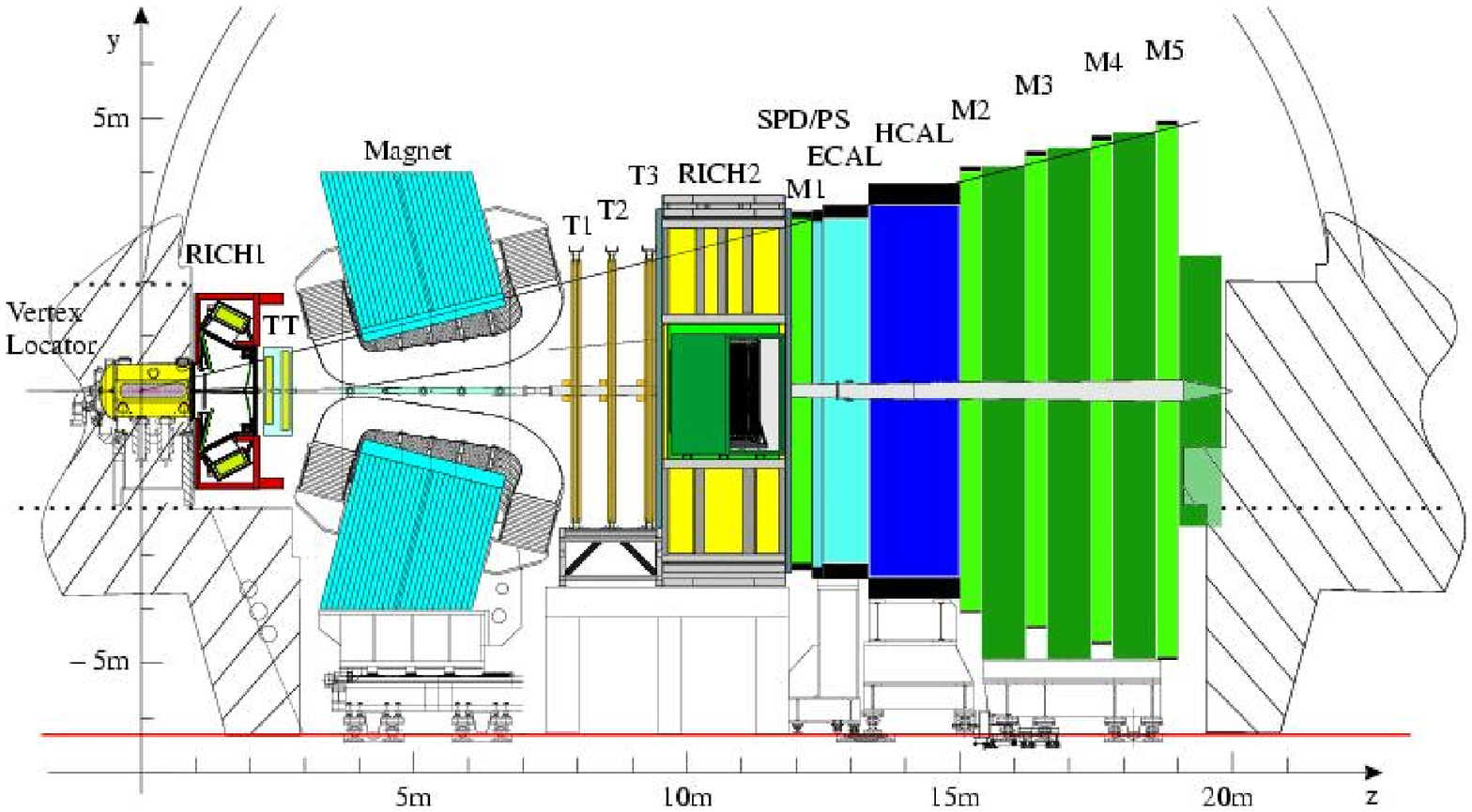} 
\caption[The \lhcb\ Detector]{%
\captionFont The \lhcb\ Detector\label{fig:det.lhcb.detectorx}}
\end{center}
\end{figure}
 \lhcb\ is specifically designed to make best use of the large number
 of \bbar\ pairs produced at the \lhc. The \lhcb\ detector is a single
 arm spectrometer with an angular acceptance from an outer limit of
 \un{250}{mrad} in the non-bending plane, and \un{300}{mrad} in the
 bending plane, down \un{15}{mrad}. This geometry is motivated by the
 kinematics of \bbar\ production in high energy proton--proton
 collisions, as discussed above.

 Amongst the most important features of the the \lhcb\ detector are:
\begin{itemize}
\item Acceptance down to small polar angles / large pseudo-rapidity,
      to maximise B-hadron yield.
\item Excellent proper time resolution to exploit the full \prt{B}
 physics potential at the \lhc, including measurements in the rapidly
 oscillating \Bso\ system.
\item Particle identification by two Ring Imaging CHerenkov (RICH)
 counters, for clean data samples and flavour tagging with Kaons.
\item Dedicated \prt{B} trigger, including high \pt\ hadron and
 lifetime triggers for high efficiency.
\end{itemize}
 Figure \ref{fig:det.lhcb.detectorx} shows a schematic overview of
 the \lhcb\ detector.  It comprises a vertex detector system, which
 includes the pile--up veto counter; a magnet and a tracking system;
 two RICH counters; an electromagnetic calorimeter and a hadron
 calorimeter, and a muon detector. All detector sub--systems, except
 for RICH~1, are split into two halves that can be separated
 horizontally for maintenance and access to the beam pipe.

\subsubsection{Material Budget}
 LHCb has recently undergone a major re-optimisation \cite{reotdr},
 which led to a substantially reduced material budget. Particular
 weight reduction has been achieved in the following subsystems:
\begin{itemize}
  \setlength{\itemsep}{2ex plus3ex minus3ex}
  \item Beam pipe: Now made from Be or Al/Be alloy.
  \item Vertex Detector: 21, \un{220}{\mu m} thin detector elements.
  \item RICH: Mirrors are now to be made from light materials (either
        Carbon fibre or Beryllium), support structures have been moved
        outside the acceptance.
  \item Tracking: All tracking stations inside the magnet have been
        removed. There is 1 (double) station before, and 3 stations behind
        magnet.
\end{itemize}
 The resulting material ``seen'' by a particle before RICH~1 is
typically $40\%$ of a radiation length, and $12\%$ of an interaction
length.

\subsection{Magnet}
 To achieve a precision on momentum measurements of better than half
 a percent for momenta up to \un{200}{GeV}, the LHCb dipole provides
 integrated field of \un{4}{Tm}. As seen in figure
 \ref{fig:det.lhcb.detectorx}, the magnet poles are inclined to follow
 the LHCb acceptance angles. This allows the \un{4}{Tm} to be retained
 with a power consumption of \un{4.2}{MW}. The warm magnet design
 chosen of LHCb allows for regular field inversions to reduce
 systematic errors in CP violation measurements. The LHCb magnet is
 currently being installed in the collision hall.

\subsection{Tracking}
 The LHCb tracking system consists of the Vertex Locator (VELO), one
 tracking station before the magnet (``Trigger Tracker''), and three
 tracking stations behind the magnet (T1 - T3).
 The VELO, the Trigger Tracker, and high-occupancy regions in T1-T3
 near the beam line (``Inner Tracker'') use Si technology, while the
 outer regions in T1-T3 (``Outer Tracker'') use straw tube drift
 chambers.

\subsubsection{The Vertex Locator}
\label{sec:VELO}
\begin{figure}
\begin{center}
\begin{tabular}{ccc}
\parbox{0.33\textwidth}{
\includegraphics[width=0.38\textwidth]{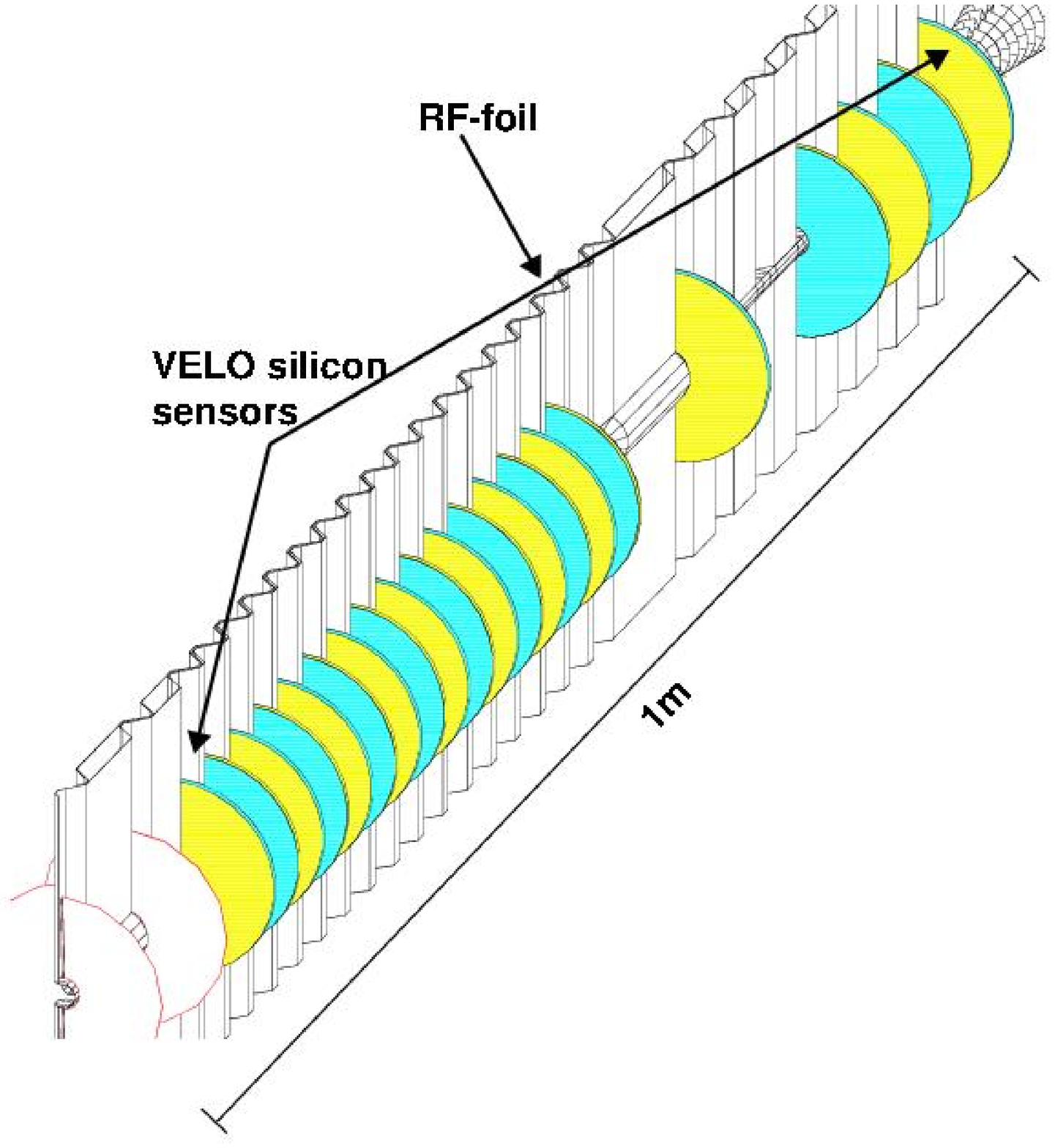}}
&
\parbox{0.33\textwidth}{
\includegraphics[width=0.33\textwidth]{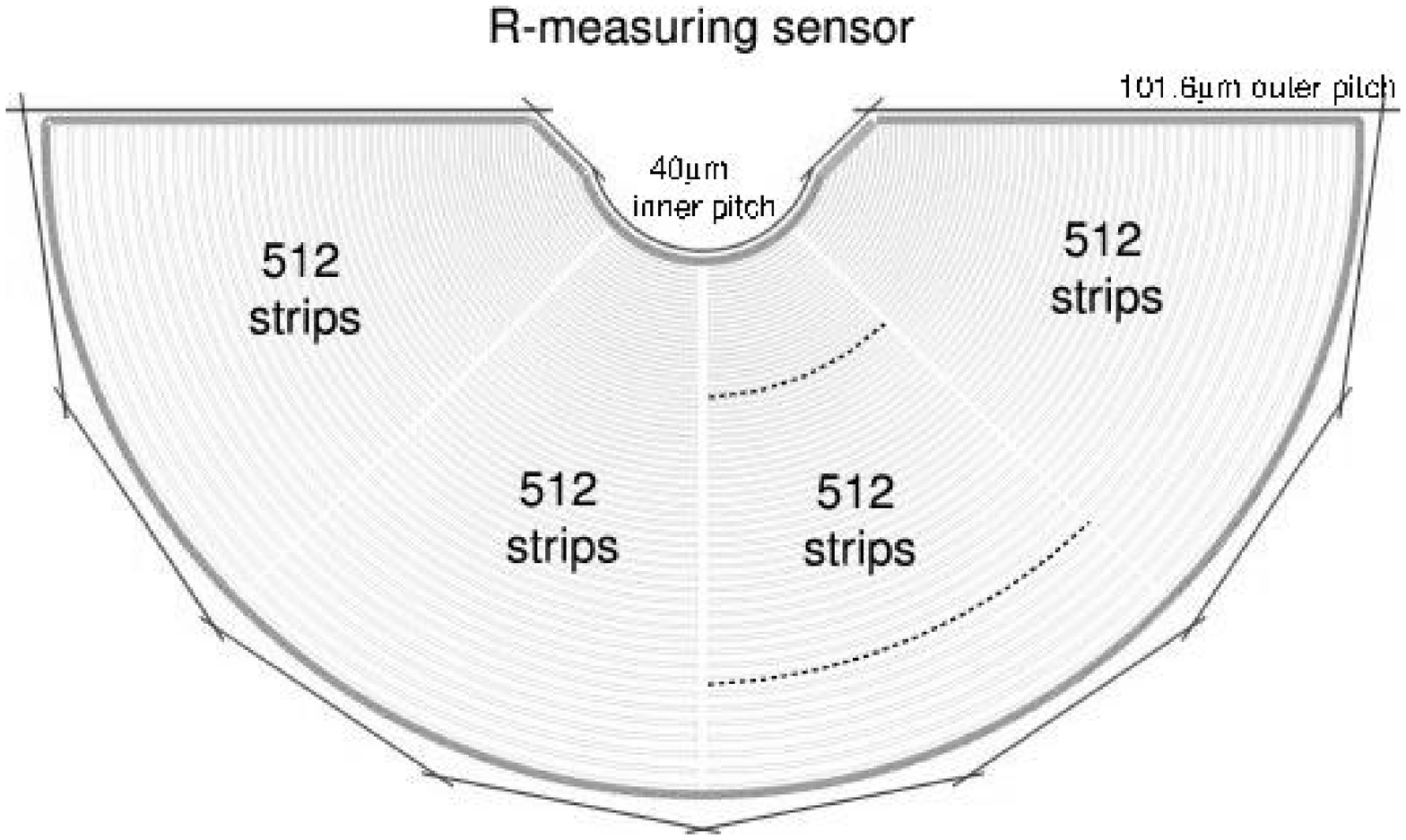}
\vspace{1ex}\\
\includegraphics[width=0.33\textwidth]{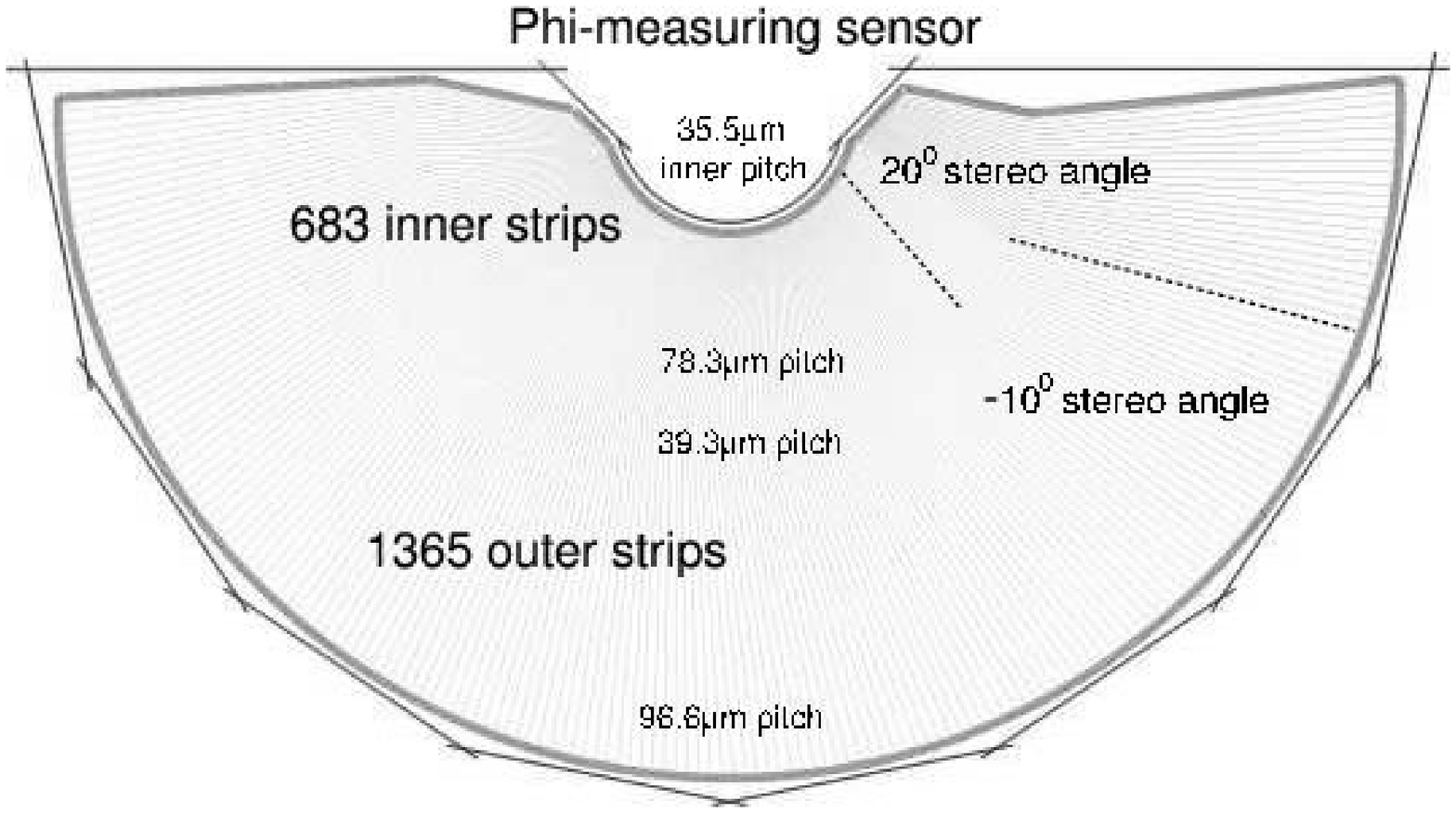}
}
&
\parbox{0.27\textwidth}{
\rotatebox{90}{\includegraphics[height=0.22\textwidth]{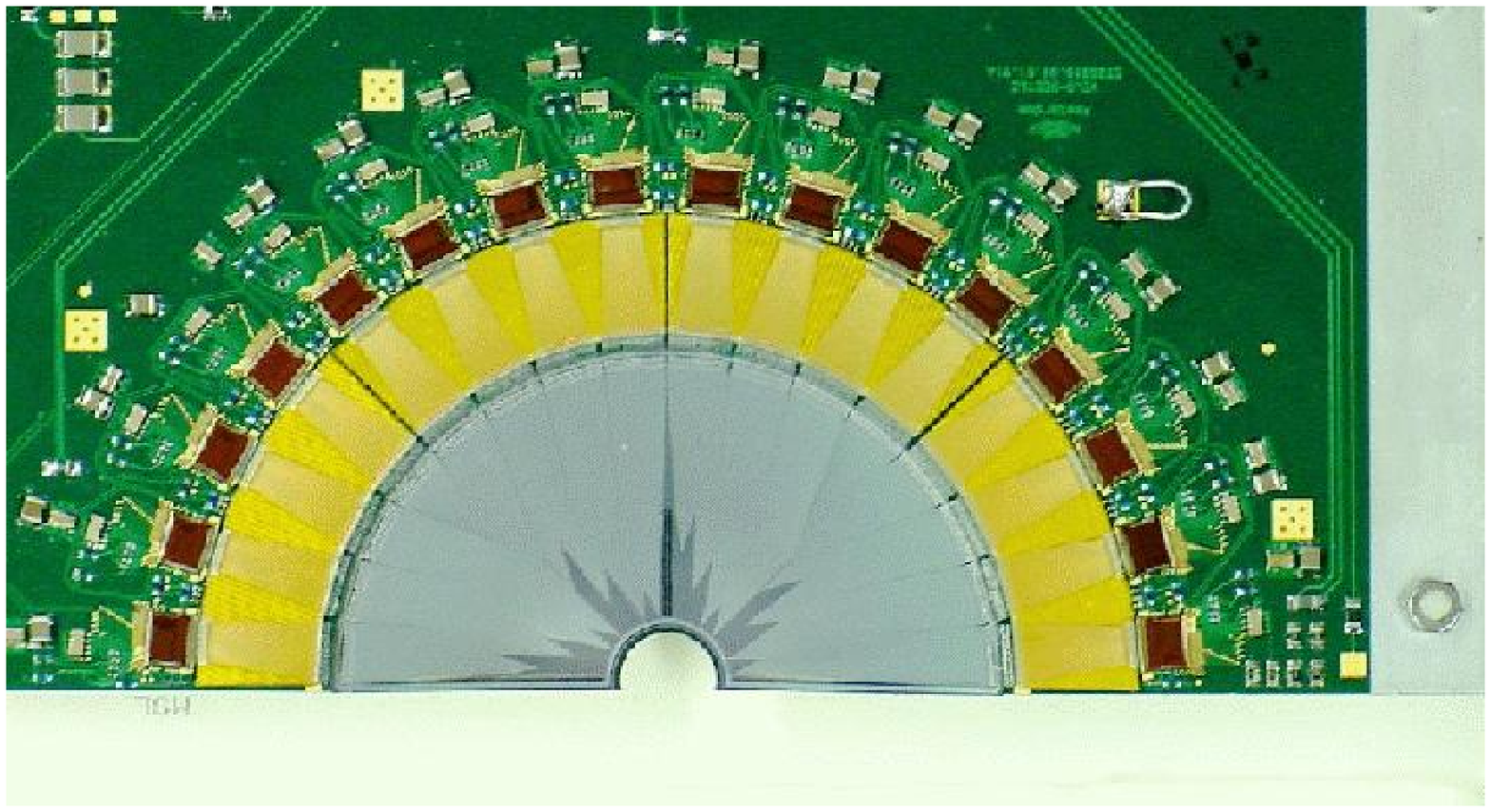}}
}
\\
\parbox[t]{0.33\textwidth}{
(a) VELO with RF-foil,  21 $r-\phi$ detector stations, and two
upstream $r$ stations for the Pile Up system.
}
&
\parbox[t]{0.33\textwidth}{(b) $r$ and $\phi$ sensors. For each sensor, 2 readout strips are
indicated by dotted lines, for illustration.}
&
\parbox[t]{0.27\textwidth}{
(c) Prototype Si sensor with readout electronics}
\end{tabular}
\caption{\captionFont The LHCb Vertex Locator (VELO)
\label{fig:Velo}}
\end{center}
\end{figure}
 To measure the time dependent decay rate asymmetries, a detector with
 excellent spatial resolution is required, especially for measurements
 in the rapidly oscillating \Bso\ system. At LHCb, this is provided by
 the Vertex Locator (VELO, Fig \ref{fig:Velo}), comprising a series of
 21 detector station placed along the beam line covering a distance of
 about \un{1}{m}. Each station consists of two pairs of half-circular
 Si microstrip detectors (Fig \ref{fig:Velo}), one pair measuring the
 radial ($r$), and one the azimuthal ($\phi$) co-ordinate.  The
 sensors are made from \un{220}{\mu m} thin Silicon, and have a
 readout pitch between \un{37}{\mu m} and \un{102}{\mu m}.  To achieve
 the required high acceptance at small polar angles (see section
 \ref{sec:bottomsAtLHC}), the sensitive area starts at only \un{8}{mm}
 from the beam line. To protect the detectors during beam injection,
 they can be retracted from the beam line. To minimise the material
 between the interaction region and the detector, the Si sensors are
 placed inside a secondary vacuum, separated from the primary vacuum
 by a \un{\sim \frac{1}{4}}{mm} thin Al foil, which also shields the
 sensors from RF pickup from the beam.  In addition to the 21 VELO
 stations which are mostly ``downstream'' of the interaction point
 (between the interaction point and the rest of the detector), there
 are two $r$-disks upstream of the interaction point which make up the
 Pile-Up System, used in the trigger \lzero\ for identifying
 multiple interaction events.
 The VELO provides an impact parameter resolution of \( \sigma_{IP} =
\un{14}{\mu m} + \frac{\un{35}{\mu m}}{\units{p_T/GeV}} \) and a time
resolution of $\sigma_{tau} \sim \un{40}{ps}$ (for \prt{\Bso \to D_s
\pi}), sufficient to resolve \prt{\Bso} oscillations up to $\Delta m_s
= \un{68}{ ps^{-1}}$ corresponding to $x_s = 105$.

\subsubsection{Tracking}
 Each tracking station consists of 4 layers. The outer layers (1 and
 4) measure the track coordinate in the bending plane
 (``$x$-layers''). The inner layers (2 and 3) are rotated by
 \degrees{+5} and \degrees{-5} respectively relative to the $x$-layers
 (``stereo layers''). This geometry optimises the
 resolution in the bending plane, for precise momentum measurements,
 while providing sufficient resolution in the non-bending plane for
 effective 3-D pattern recognition.

 The 4 layers of the Trigger Tracker are split into two sub stations,
 separated by \un{30}{cm}. This allows a rough momentum estimation
 from the bending of the tracks in the magnetic fringe field. This
 momentum information is used in the trigger \lone\ decision.
 The Si detectors in the Trigger Tracker and Inner Tracker have a read
 out pitch of \un{198}{\mu}, with strip lenghs of up to \un{33}{cm} in
 the Trigger Tracker, and \un{11-22}{cm} in the Inner Tracker. The
 thickness of Si layers is \un{320}{\mu} for the Inner Tracker, and
 \un{500}{\mu m} for the Trigger Tracker. The Outer Tracker is made of
 $\un{5}{mm}\times\un{4.7}{m}$ straw tubes, with a fast drift gas
 ($75\% \mathrm{Ar}$, $15\% \mathrm{CF_4}$, $10\% \mathrm{CO_2}$),
 allowing signal collection in less than \un{50}{ns}.
 The LHCb tracking system provides a momentum resolution of \(
 \frac{\delta p}{p} = 0.37\% \). For the example of \prt{B_s \to D_s
 K}, this translates into a mass resolution of \un{14}{MeV}.  The
 track-finding efficiency is $94\%$ for tracks with hits in all
 tracking stations. The Ghost rate is $9\%$ ($3\%$ for tracks with
 $p_T > \un{0.5}{GeV}$).

\subsection{Calorimetry} 
 The main design constraints for the calorimeter system come from its
 central role in the \lzero\ trigger decision, which must be provided
 and processed within the \un{25}{ns} between each bunch crossing.
 The general structure of the calorimeter system is as follows: the
 first element seen by a particle coming from the interaction point is
 a scintillator pad detector (SPD), that signals charged
 particles. This is followed by a \un{12}{mm} lead wall and another
 SPD, which together form the preshower detector (PS). This is then
 followed by $25$ radiation lengths ($1.1$ interaction lengths) of a 
 \chem{Pb}/scintillator Shashlik calorimeter (ECAL) and $5.6$ interaction length of
 a iron/scintillator tile hadron calorimeter (HCAL).

 Most ECAL modules, and a large fraction of HCAL modules have already
 been delivered to CERN. In testbeams an energy resolution of
\[
  \left(\frac{\sigma_{E}}{E}\right)_{\mathrm{ECAL}} 
    =         \frac{9.4\%}{\sqrt{\un{E}{/GeV}}} 
       \oplus 0.83\% 
       \oplus \frac{\un{0.145}{GeV}}{E}
,\;\;\;\;\;\;
 \left(\frac{\sigma_{E}}{E}\right)_{\mathrm{HCAL}} 
    =         \frac{75\%}{\sqrt{\un{E}{/GeV}}} 
       \oplus 10\%
\]
 has been achieved.

\subsection{Muon System}
 The muon system provides offline muon identification, and information
 for the trigger \lzero. It consists of four stations behind the
 calorimeters (M2-M5), and one unshielded station in front of the
 calorimeters (M1). Most of the muon chambers will be Multi-Wire
 Proportional Chambers (MWPC). For the central region of M1
 (\un{0.6}{m^2}), triple GEM technology will be more suitable due the
 expected high rates in that area.

\subsection{RICH}
 \lhcb\ intends to perform high precision measurements in many
 different \prt{B} decay channels. Many interesting decay channels are
 themselves backgrounds to topologically similar ones. Typically the
 branching ratios are of the order of $\sim 10^{-5}$. The particle
 identification and in particular \prt{K/\pi} separation provided by
 the RICH is essential for obtaining the clean samples needed to
 perform a comprehensive range of high--precision \cp\ violation
 measurements, and allows the use of Kaons of flavour tagging,
 dramatically improving the tagging performance at LHCb
 \ref{sec:bTagging}.
\begin{figure}
\begin{center}
\begin{tabular}{cccc}
 RICH~1 & Rings in RICH~1 & RICH~2 (with     & Rings in RICH~2 \\
        &                 & RICH~1 for scale)& 
 \\
 \includegraphics[height=0.25\textwidth]{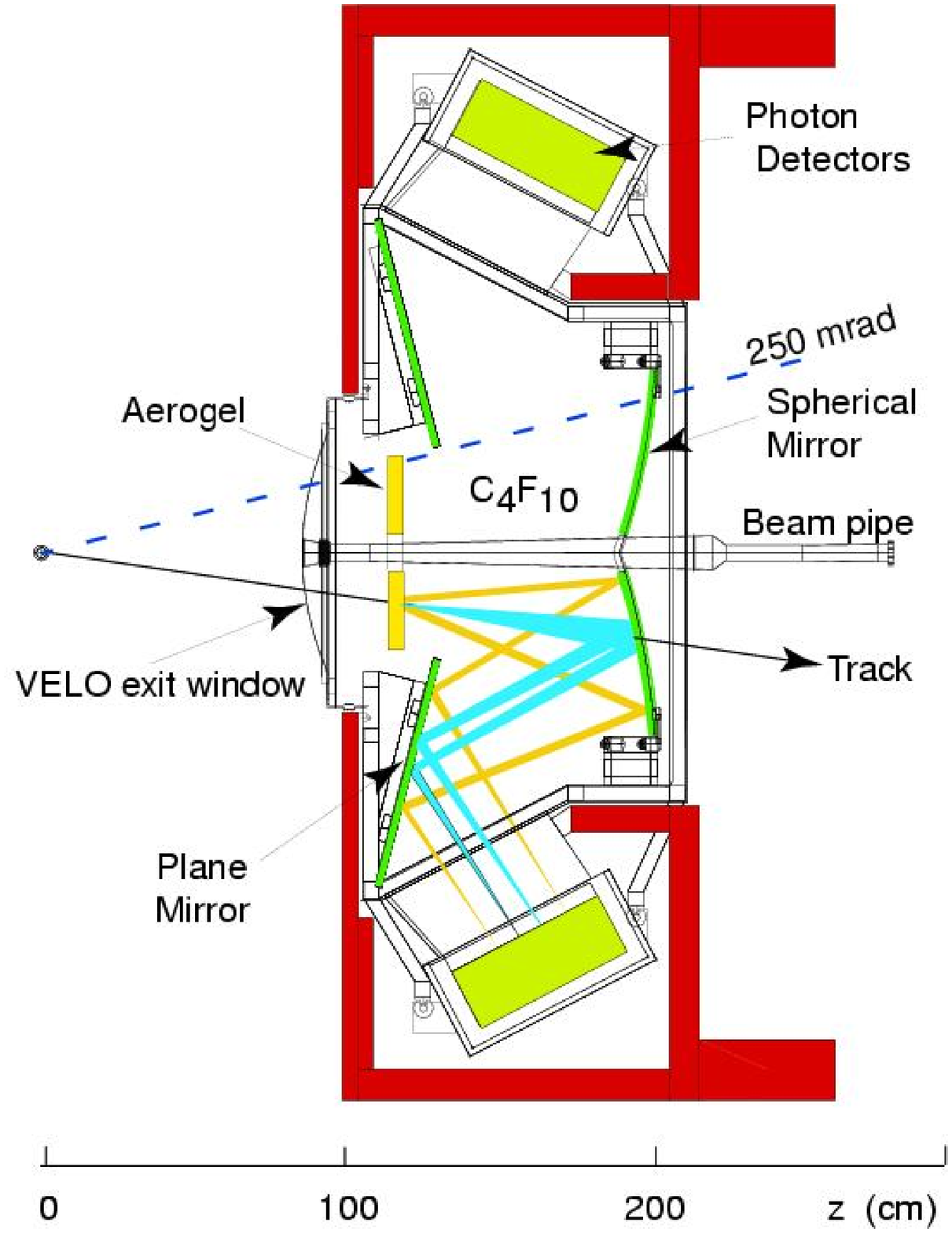}
&
 \includegraphics[height=0.25\textwidth]{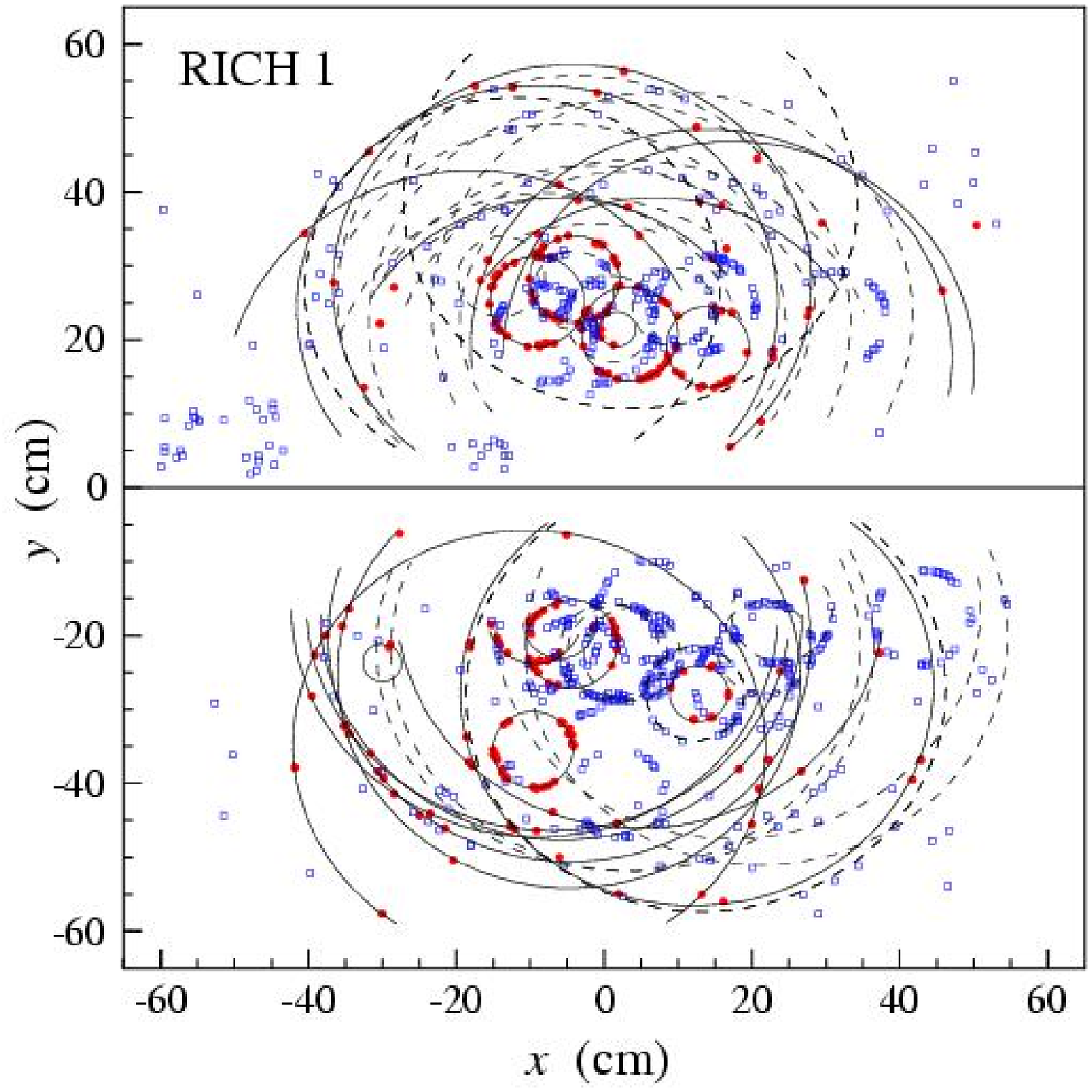}
&
 \includegraphics[height=0.25\textwidth]{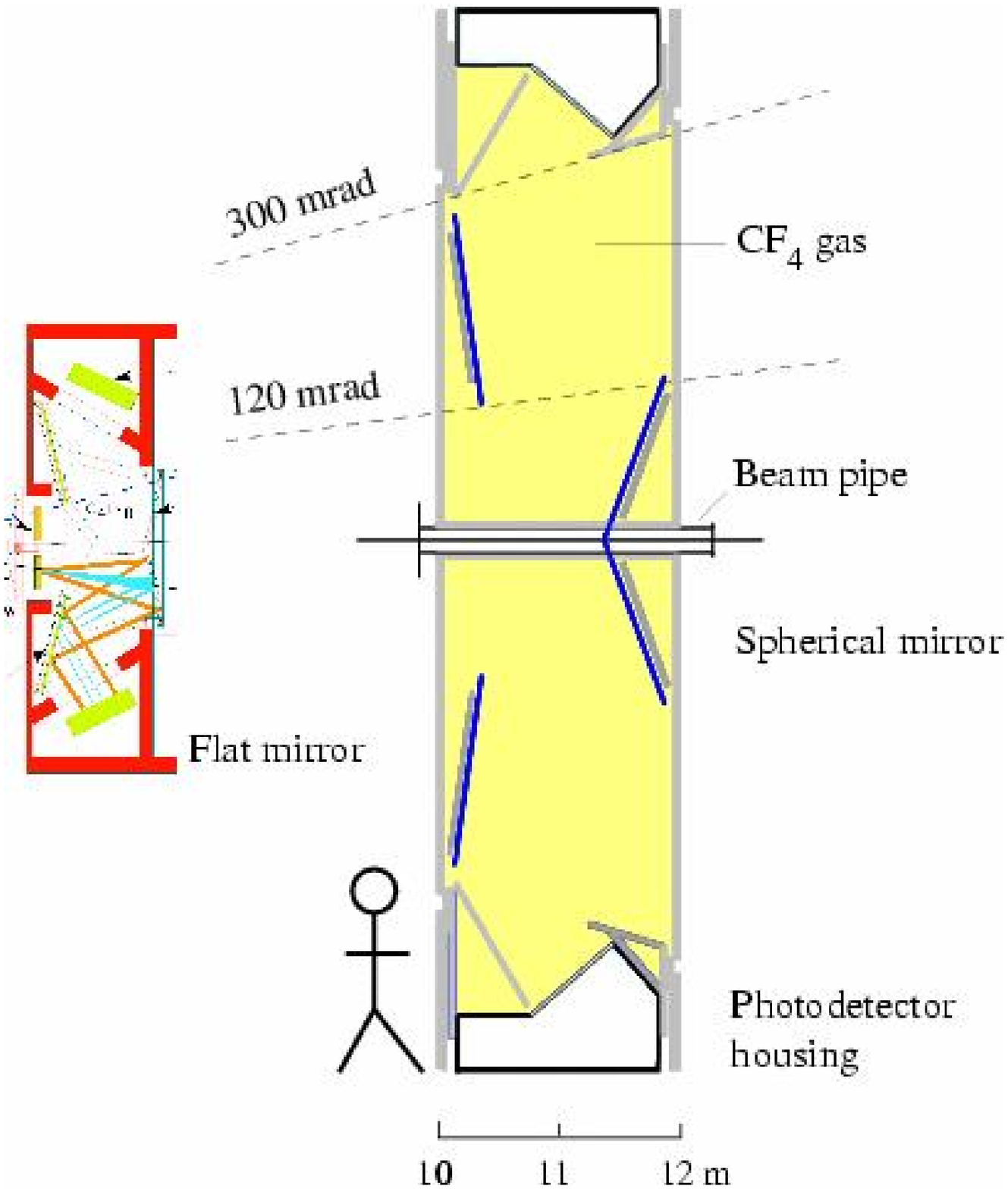}
&
 \includegraphics[height=0.25\textwidth]{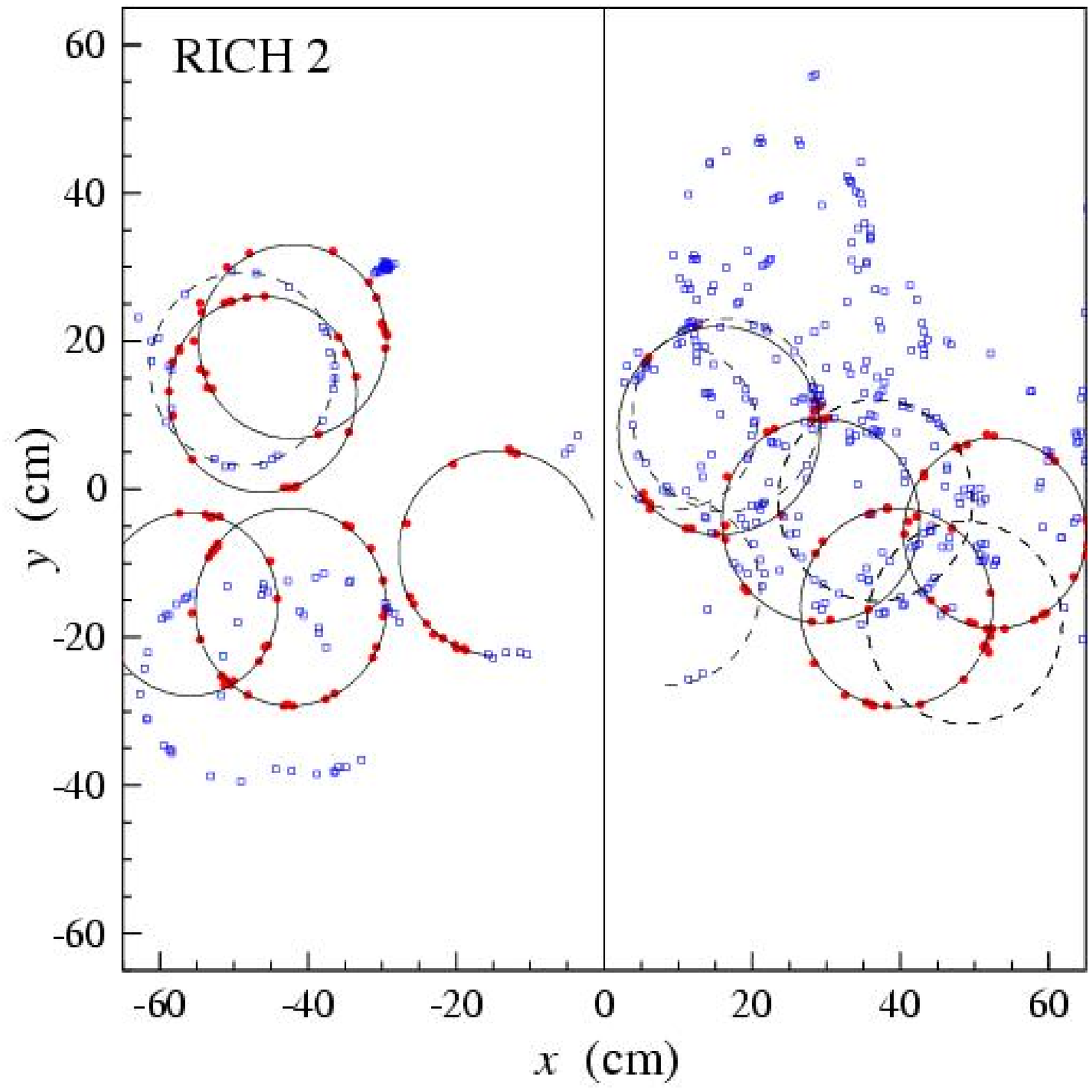}
\end{tabular}
\caption{\captionFont The LHCb RICH system used 2 detectors and 3 radiators, to
  provide \prt{K-\pi} separation from \un{\sim 1}{GeV} to beyond
  \un{100}{GeV}. The figure shows RICH~1 and RICH~2 with the
  respective event displays for a typical event. The points represent
  the detector response. The rings drawn through the points are the
  result of a global pattern recognition algorithm, which uses the
  tracking results as a seed.\label{fig:riches}}
\end{center}
\end{figure}

\begin{figure}
\begin{center}
\begin{tabular}{ccc}
\parbox{0.32\textwidth}{ (a) Momenta of tracks in \prt{B\to \pi\pi}
 events, and tagging Kaons. For both, \prt{K/\pi} separation is
 essential.}
&
\parbox{0.3\textwidth}{
 (b) \C\ angle $\theta_C$ vs momentum for Kaons an pions for each
 radiator in the LHCb RICH.
}
&
\parbox{0.3\textwidth}{
 (c) Polar angle $\theta$ vs momentum
     for all tracks in $B_d^0\to \pi\pi$ 
     evts, approx. RICH-coverage indicated.
}
\\
\parbox{0.32\textwidth}{
\includegraphics[height=0.29\textwidth]{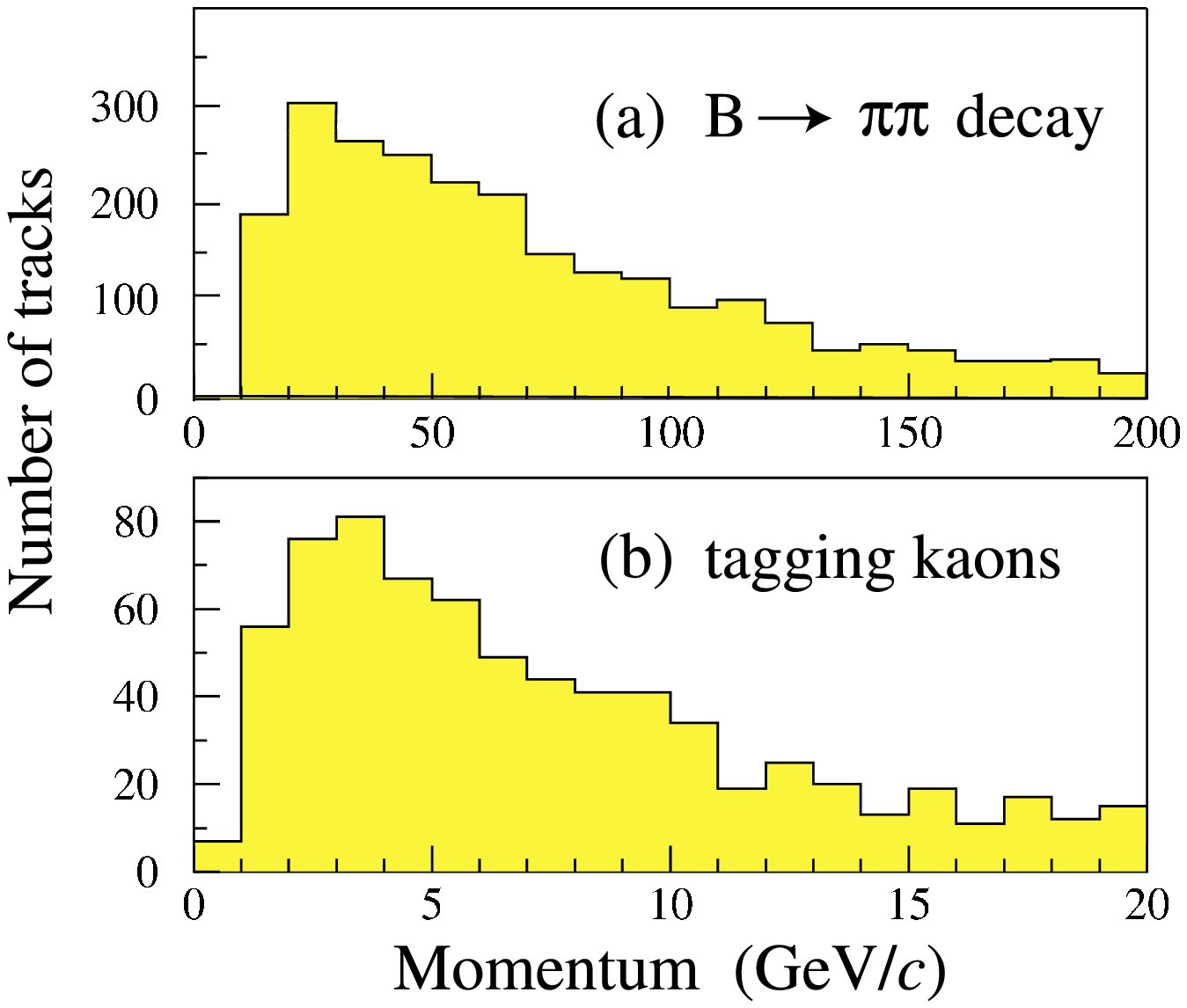}}
&
\parbox{0.3\textwidth}{
{\small
\parbox[t]{1ex}{\mbox{}\vspace{-0.0ex}\\\rotatebox{90}{$\theta_C$ \units{[mrad]}}}
\hspace{0.2ex}
\parbox[t]{0.275\textwidth}{\mbox{}\\
\includegraphics[height=0.275\textwidth]{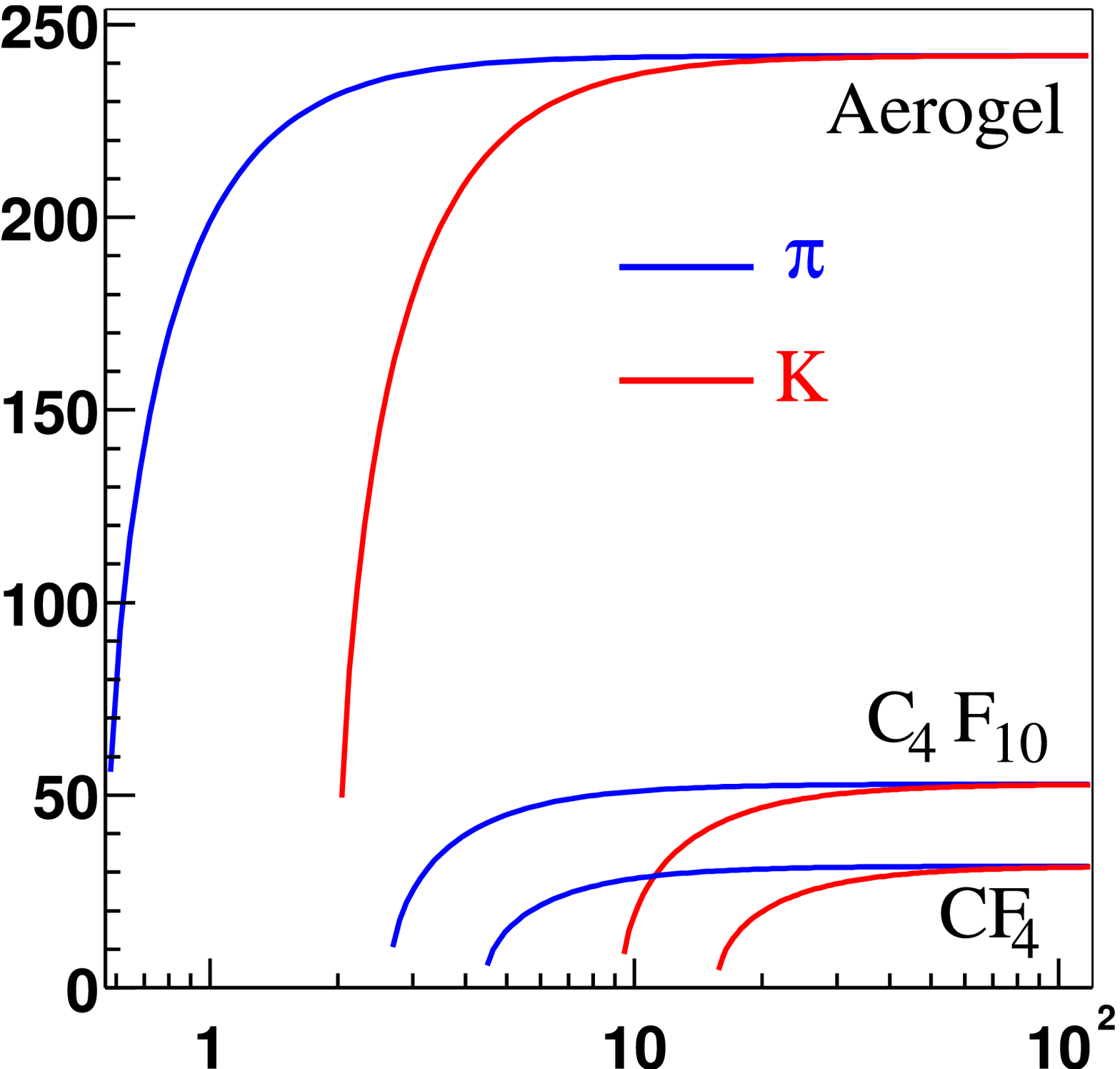}
\\\mbox{}\hfill $p$ \units{[GeV]}
}}}
&
\parbox{0.3\textwidth}{
\includegraphics[height=0.3\textwidth]{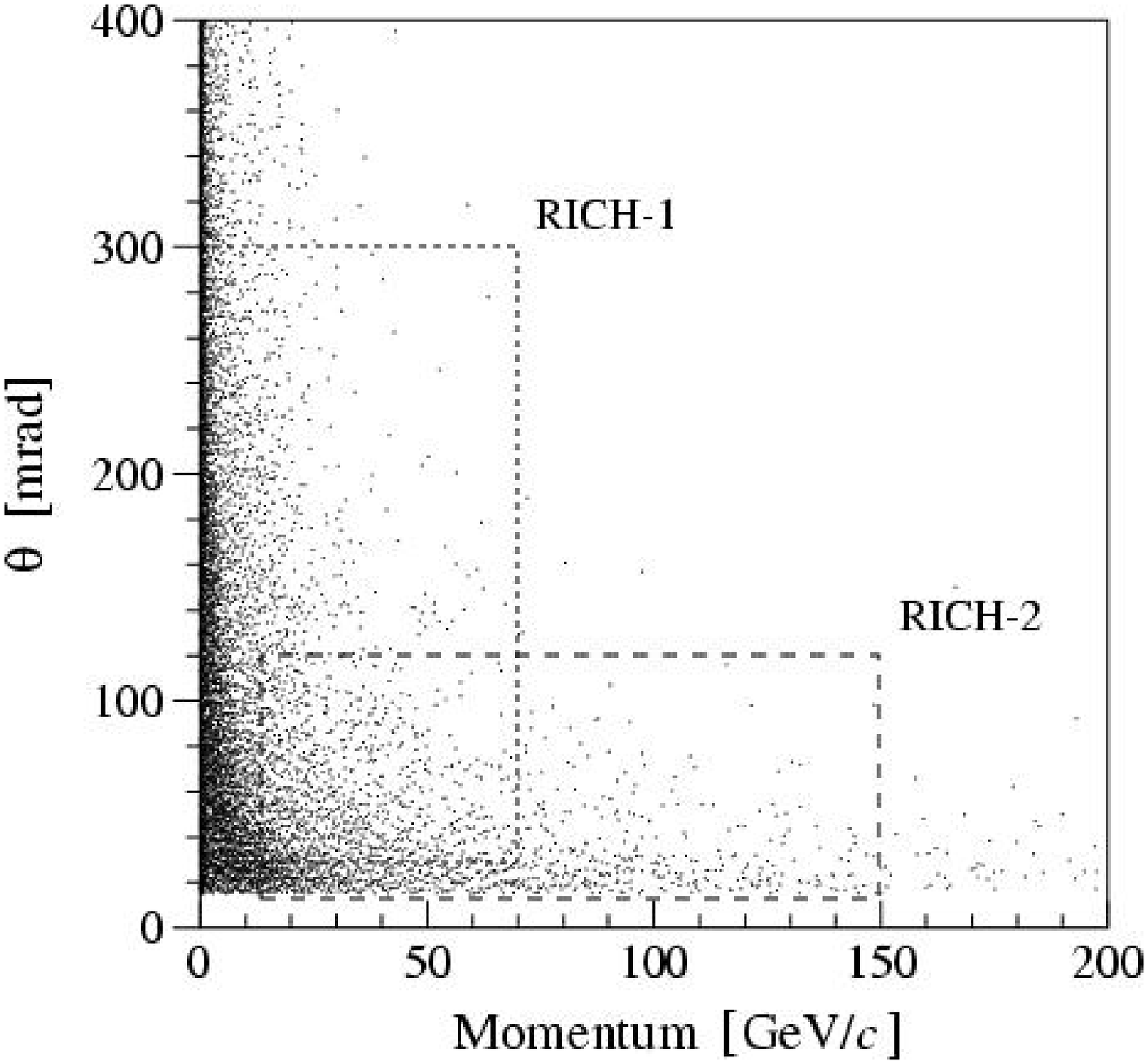}}
\end{tabular}
\caption{\captionFont The LHCb RICH employs 3 Radiators in 2 RICH detectors to
  provide \prt{K/\pi} separation from \un{\sim 1}{GeV} to beyond
  \un{100}{GeV}.\label{fig:richMomenta}}
\end{center}
\end{figure}
 Figure \ref{fig:richMomenta} (a) shows the momentum distribution of
 (a) pions in \Bpipi\ events, and (b) tagging Kaons. This illustrates
 the need for \prt{K/\pi} separation over a wide range of momenta;
 \lhcb\ seeks \prt{K/\pi} separation from momenta of \un{\sim 1}{GeV}
 to beyond \un{100}{GeV}.

 RICH (Ring Imaging CHerenkov) counters measure the opening angle
 $\theta_C$ of the \C\ cone emitted by particles as they traverse a
 transparent medium, by imaging it onto an array of photo detectors as
 illustrated in Fig \ref{fig:riches}. This opening angle depends on
 the speed of the particle. Combining it with the momentum information
 from the tracking system, allows to identify the particle by its
 mass. To cover a momentum range from \un{\sim 1}{GeV} to beyond
 \un{100}{GeV}, LHCb employs two RICH detectors and three radiators,
 Aerogel ($n=1.03$) and \chem{C_4 F_{10}} gas ($n=1.0014$) in RICH~1
 and \chem{CF_4} gas ($n=1.0005$) in RICH~2.  The angular and
 approximate momentum coverage of the two RICH detectors at LHCb is
 shown in \ref{fig:richMomenta}, superimposed over a scatter plot
 showing polar angles and momenta of particles in \prt{B_d \to \pi\pi}
 events.
\begin{figure}
\begin{center}
\begin{tabular}{ccc}
\multicolumn{2}{c}{(b) Inv. mass of
                        \prt{B_s\to D_s K}} with \prt{B_s \to D_s \pi}
                        background\\
                        & without RICH   &  with RICH 
\\
 \includegraphics[width=0.32\textwidth]{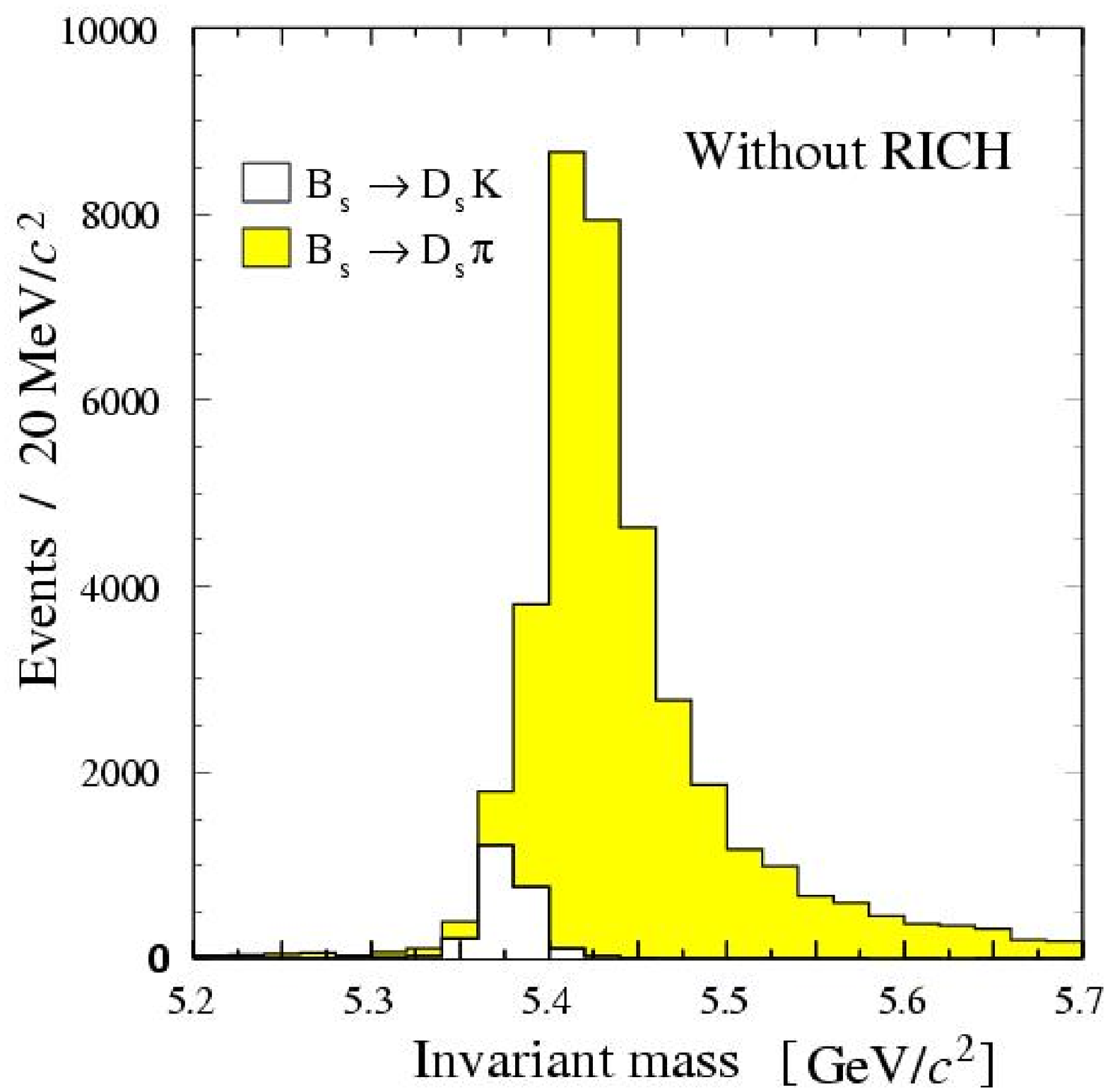}
&
 \includegraphics[width=0.32\textwidth]{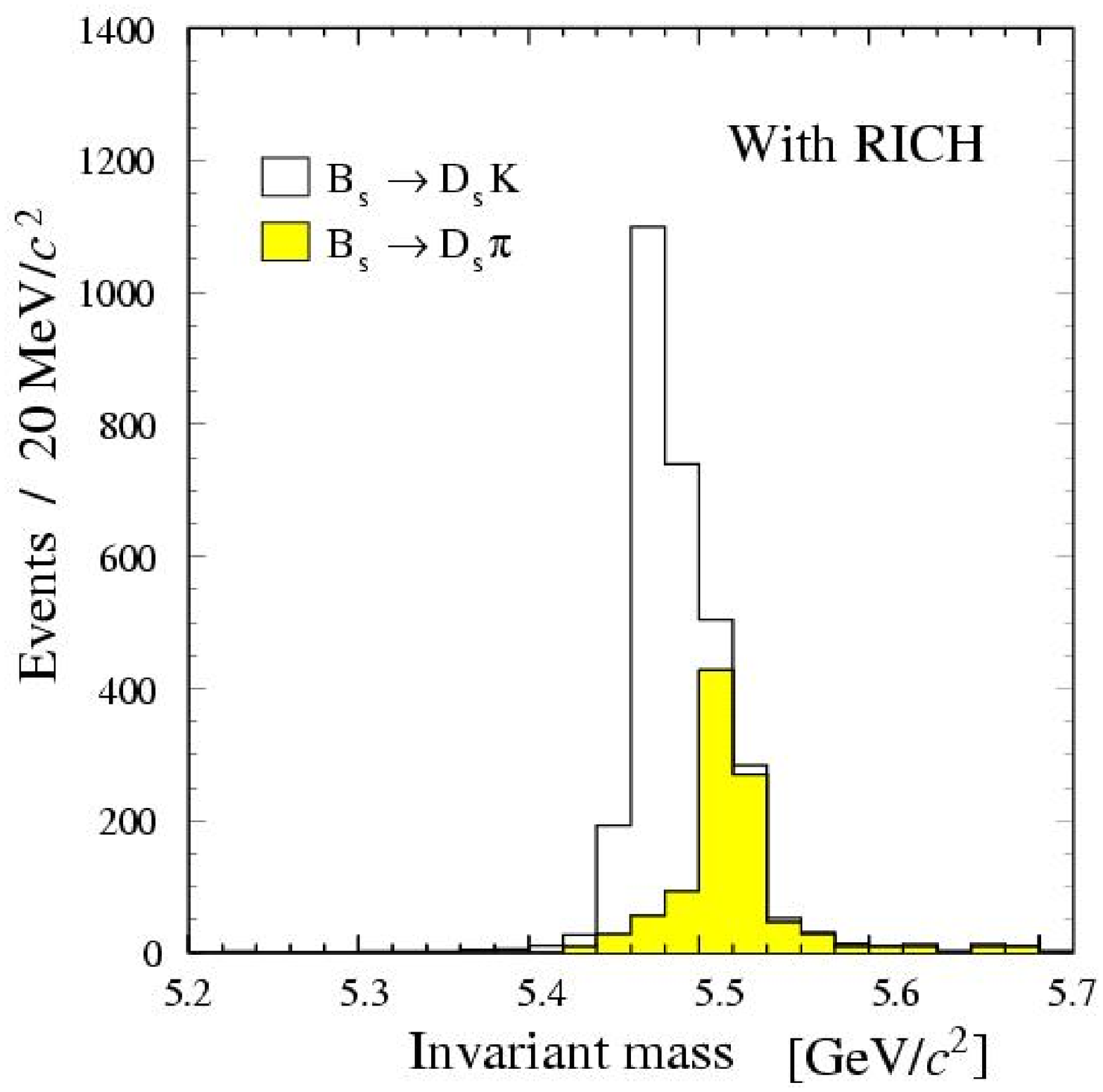}
\end{tabular}
\end{center}
\caption{\captionFont Without RICH particle ID, many decays
  would be swamped by topologically similar background. Illustrated
  here for the $\gamma$-sensitive channel \prt{B_s\to D_s K}, which
  has a $\sim 10$ times smaller B.R. than its dominant background,
  \prt{B_s\to D_s \pi}.\label{fig:richPerf}}
\end{figure}
 Figure \ref{fig:richPerf} illustrates how the RICH particle ID
 cleans up the \prt{B_s\to D_s K} signal, a $\gamma$-sensitive channel
 that would otherwise be completely dominated by background from
 \prt{B_s\to D_s \pi} which has a $\sim 10$ times higher branching
 fraction.

\subsection{LHCb Trigger}
\label{sec:trigger}
 The LHCb trigger has the task of reducing the event rate of
 \un{40}{MHz} by a factor of $200,000$ to the write-to-tape rate of
 \un{200}{Hz}, while keeping as many interesting B-events as
 possible. This is achieved in three steps.
\begin{itemize}
 \item \lzero\ uses information from the Pile-Up detector, the
  Calorimeters and the Muon Chambers, to reduce the event rate from
  \un{40}{MHz} to \un{1}{MHz}.
 \item \lone\, uses momentum and impact parameter information
 from the VELO and the Trigger Tracker, to reduce the event rate
 further to \un{40}{kHz}.
 \item The High Level Trigger (\hlt) will have access to the complete
 event information to perform full event reconstruction.
\end{itemize}
 While the \lzero\ algorithm will run on dedicated hardware, \lone\
 and the \hlt\ will run trigger software on computing farms built from
 off-the-shelf components.

\subsection{Flavour Tagging}
\label{sec:bTagging}
 To measure the time dependent decay rate asymmetries from which the
 CKM phases are extracted, the flavour of the reconstructed B meson at
 the time of creation needs to be known. Usually, this is done by
 looking at B decay products from the
 opposite-side B-hadron\footnote{%
Note that the ``opposite side'' B hadron
usually travels into a similar direction as the B hadron of interest,
which is crucial given LHCb's detector geometry
}
 created alongside the one being reconstructed.  (``lepton tag'',
 ``Kaon tag'', ``Vertex Charge'').
 
 An alternative strategy is same side tagging, which uses the
 correlation between the flavour of a \prt{B_s} meson and the charge
 of a \prt{K^+} picking up the 2\nd\ \qrk{s} quark produced in the
 process.  In principle, this method also works with \prt{B_d} mesons
 and pions, but given the large number of pions created in a hadron
 collider, it is much more difficult to pick out the right one.

\begin{table}
\begin{center}
\begin{tabular}{cc}
\parbox{0.45\textwidth}{
\begin{tabular}{|c|c|c|c|}
\hline
Tag & $\varepsilon$
    & $\omega$ 
    & $\varepsilon_{\mathrm{eff}}$ 
\\\hline
\prt{\mu} & 11\% & 35\% & 1\%
\\
\prt{e} & 5\% & 36\% & 0.4\%
\\
\prt{K_{\mathrm{opp-side}}} & 17\% & 31\% & 2.4\%
\\
$Q_{\mathrm{Vtx}}$ & 14\% & 40\% & 1\%
\\
\bf \prt{B^0_d} all & \bf 41\% & \bf 35\% & \bf 4\%
\\\hline
\prt{K_{\mathrm{same-side}}} & 18\% & 33\% & 2.1\%
\\
\bf \prt{B^0_s} all & \bf 50\% & \bf 33\% & \bf 6\%
\\\hline
\end{tabular}
}
&
\parbox{0.45\textwidth}{
 The figure of merit for the tagging performance is given by the
 ``effective tagging efficiency'' $\varepsilon_{\mathrm{eff}}$ (also
 known as $\epsilon D^2$):
\(
 \varepsilon_{\mathrm{eff}} = \varepsilon \left(1-\omega\right)^2
\)
 The statistical
 significance of $N$ events with an effective tagging efficiency
 $\varepsilon_{\mathrm{eff}}$ is equivalent to
 $\varepsilon_{\mathrm{eff}} N$ perfectly tagged events.
}
\end{tabular}
\caption{\captionFont Tagging efficiencies ($\varepsilon$), wrong-tag fractions ($\omega$)
and effective tagging efficiencies 
$\varepsilon_{\mathrm{eff}} \equiv \varepsilon(1-2\omega)^2$ for 
\prt{B\to hh}.\label{tab:tagging}}
\end{center}
\end{table}
 Table \ref{tab:tagging} shows the expected tagging performance at LHCb for
 \prt{B \to \pi\pi} and \prt{B_s \to KK}.

\section{Physics}
 By the year 2007, we expect a very precise measurement of the angle
 $\beta$, from the B-factories and the Tevatron, and results for the
 mass and lifetime difference of the CP eigenstates of the \prt{B_s}
 system, $\Delta m_s$ and $\Delta\Gamma_s$ from the Tevatron. With its
 huge number of \bbar\ pairs, LHCb will be able to significantly
 improve the precision on all of these measurements within the first
 year of data taking. However, here we focus on one of the most
 exciting Physics prospects at LHCb, the experiment's ability to
 perform precision measurements of the angle $\gamma$ in many
 different decay channels both in the \prt{B_d} and \prt{B_s}
 system. Both, the B-factories and CDF expect some measurement of
 $\gamma$ by 2007, but this is unlikely to be precise enough to
 provide a significant constraint on the \UT. LHCb will measure
 $\gamma$ in many different channels, some more and some less
 susceptible to \np, with a typical precision of $\degrees{5} -
 \degrees{15}$ for each channel after one year of data taking. We will
 demonstrate on three examples different strategies of measuring
 $\gamma$ at LHCb.

\subsection{\prt{B_d \to \pi\pi} and \prt{B_s \to KK}}
\subsubsection{Principle}

\begin{figure}
\begin{tabular}{*{3}{p{0.33\textwidth}}}
If there were only the tree contribution...
&
\prt{B^0_d\to \pi\pi} would measure $2(\bet + \gam)$...
&
... but there are Penguins:
\\
\includegraphics[width=0.33\textwidth]{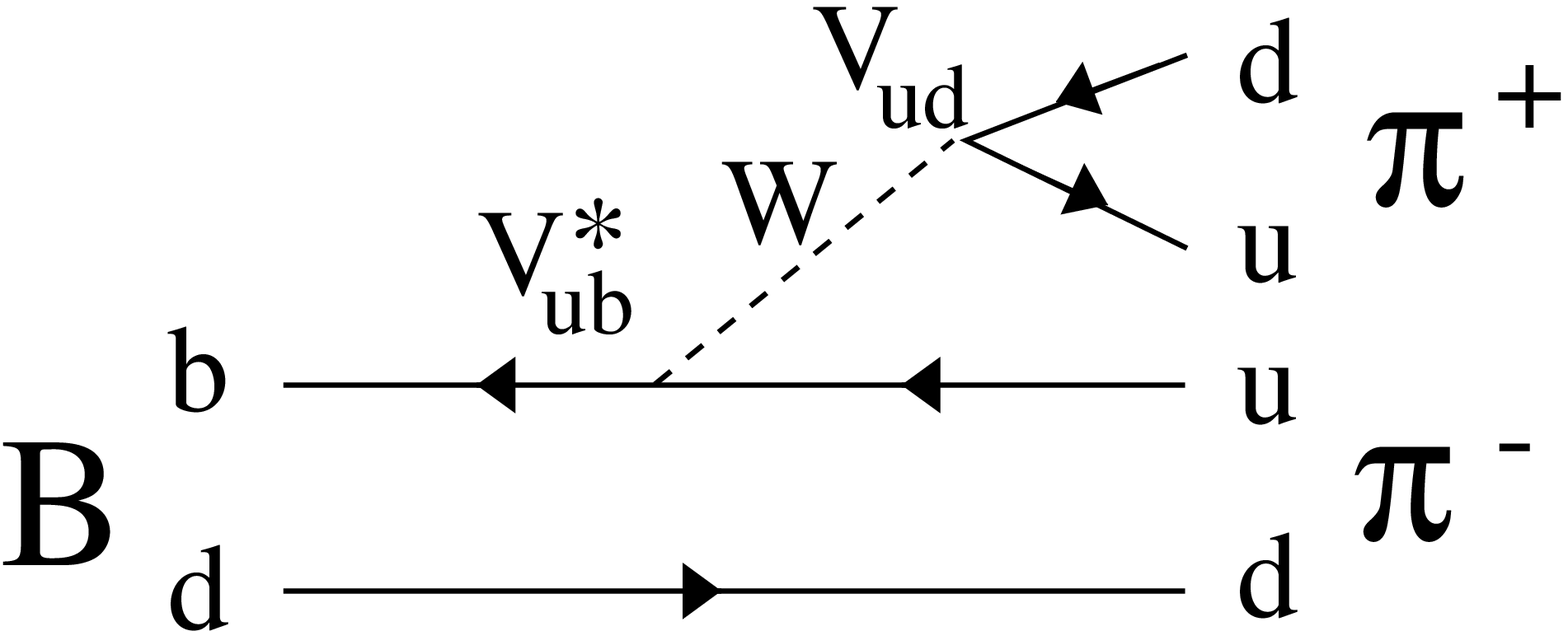}
&
\includegraphics[width=0.33\textwidth]{\slideFig/interference_pipi}
&
\includegraphics[width=0.33\textwidth]{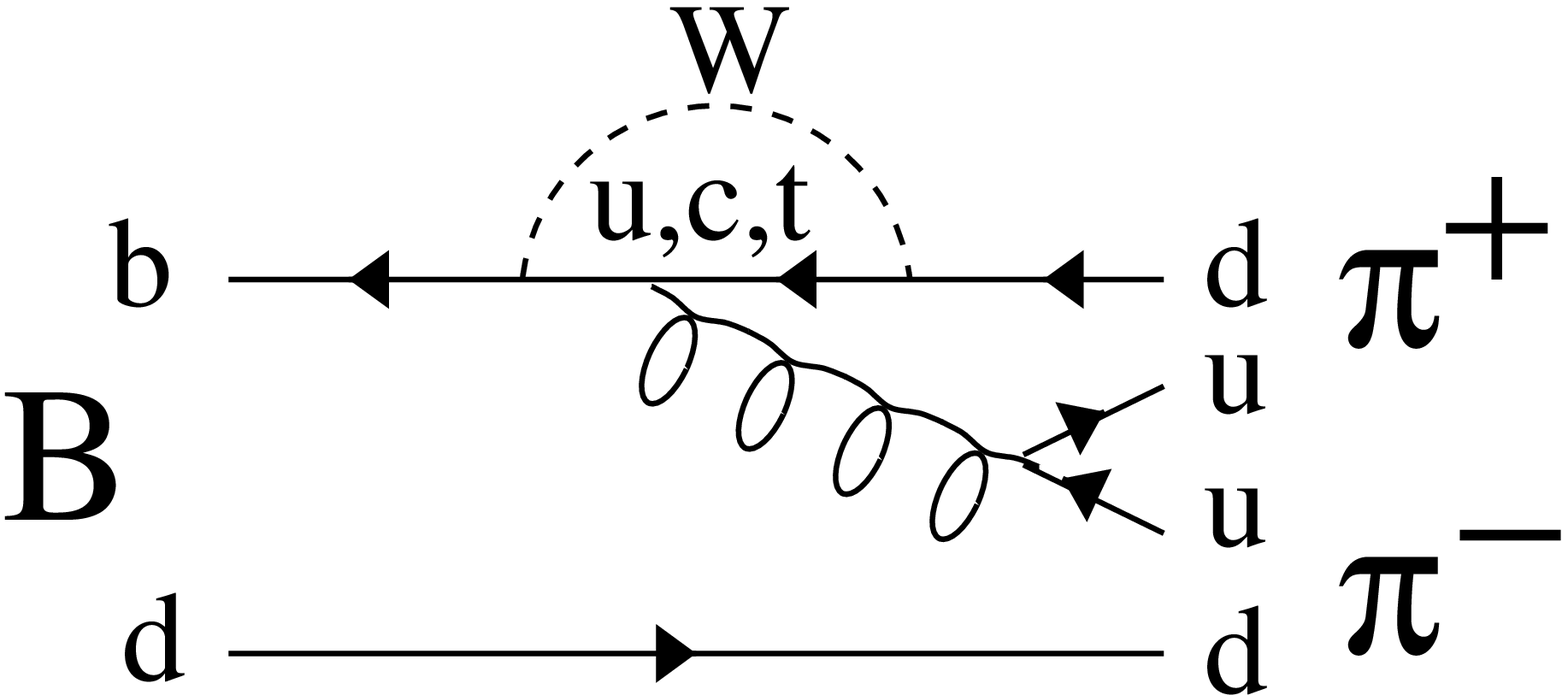}
\end{tabular}
\caption{\captionFont The phase difference between \prt{B_d \to \pi^+\pi^-} and
  \prt{B_d \to \bar{B}_d \to \pi^+\pi^-} is affected by both tree and
  penguin contributions. These can be disentangled by simultaneously
  analysing the U-Spin related decay \prt{B_s \to
  KK}.\label{fig:bpipiDiagrams}}
\end{figure}

 The decay \prt{B_d\to\pi\pi} is, due to the $b\to u$ transition in
 the tree diagram in figure \ref{fig:bpipiDiagrams}, sensitive to the
 CKM angle $\gamma$. However, the presence of penguin contributions
 severely complicates the interpretation of the observed CP asymmetries
 in terms of CKM angles. At the same time, penguin diagrams are
 interesting, because they are sensitive to \np.  A possible strategy
 that allows the tree and penguin contributions to be disentangled,
 and thus measure $\gamma$, is due to Fleischer \cite{Fleischer}, and
 uses U-spin symmetry of the strong interaction to relate observables
 in \prt{B \to \pi\pi} and \prt{B_s to KK}.
 The time dependent decay rate asymmetry can be parametrised as
\begin{equation}
 \frac{     \Gamma \left(\prt{B_d \to \pi^+\pi^-}       \right) 
          - \Gamma \left(\prt{\bar{B}_d \to \pi^+\pi^-}\right)
      }{    \Gamma \left(\prt{B_d \to \pi^+\pi^-}       \right) 
          + \Gamma \left(\prt{\bar{B}_d \to \pi^+\pi^-}\right)
      }
        = \Adirpp \cos(\Delta m_d \tau) + \Amixpp \sin(\Delta m_d
 \tau),
\end{equation}
 and similarly for \prt{B_s \to KK}. This provides four observables:
 \Adirpp, \Amixpp, \AdirKK, and \AmixKK. These can be parametrised
 with the following seven parameters:
\begin{itemize}
  \item $d, \theta$ hadronic parameters describing
        ``penguin-to-tree'' ratio and phase in \prt{B_d}.
  \item $d^{\prime}, \theta^{\prime}$ hadronic parameters related to
         ``penguin-to-tree'' ratio and phase for \prt{B_s}.
  \item $\phi_d$ = \prt{B_d} mixing phase, $2\beta$ in SM.
  \item $\phi_s$ = \prt{B_s} mixing phase, $2\delta\gamma\approx 0$ in
        SM.
  \item \gam\ is what we want to measure.
\end{itemize}
These can be reduced to three parameters, as follows
\begin{itemize}
 \item $2\beta$ and $2\delta\gamma$ will be known precisely from
 \prt{B\to J/\!\psi K_s} and \prt{B_s \to J/\!\psi \phi}
 \item $d, \theta, d', \theta'$ depend on the strong interaction
 only. Assuming U-spin symmetry, we set
   $d = d'$ and $\theta = \theta'$.
\end{itemize}
 Further details can be found in \cite{Fleischer}.
\begin{table}
\begin{center}
\begin{tabular}{||c|c|c|c|c||}
\hline\hline
 Channel & \# evts    & B/S     & tagging & mistag \\
         & per year   &         & eff     & frac   \\\hline\hline
\prt{B \to \pi\pi}
         & \un{26}{k} & $<0.7$  & $41.8\%$ & $34.9\%$
\\\hline
\prt{B_s \to KK}
         & \un{37}{k} & $<0.5$  & $49.8\%$ & $33.0\%$
\\\hline\hline
\end{tabular}
\caption{\captionFont Expected reconstruction and tagging performance for \prt{B
    \to \pi\pi} and \prt{B_s \to KK}.\label{tab:BpipiYield}}
\end{center}
\end{table}
\begin{table}[t]
\begin{center}
\begin{tabular}{|@{\,}c@{\,}|c|c|c|c|c|c|}
\cline{1-5} 
\dms & 15 & 20 & 25 & 30 & \multicolumn{2}{c}{} \\
\cline{1-5}
$\sigma(\gamma)$ & 4.0 & 4.9 & 5.9 & 8.5 & \multicolumn{2}{c}{} \\
\cline{1-5} \multicolumn{4}{c}{} \\[-1.5ex] \cline{1-4}
\DGsGs & 0 & 0.1 & 0.2 & \multicolumn{3}{c}{} \\
\cline{1-4}
$\sigma(\gamma)$ & 5.2 & 4.9 & 4.5 & \multicolumn{3}{c}{} \\
\cline{1-4} \multicolumn{4}{c}{} \\[-1.5ex] \hline
$\gamma$ & 55 & 65 & 75 & 85 & 95 & 105 \\
\hline
$\sigma(\gamma)$ & 5.8 & 4.9 & 4.3 & 4.7 & 4.7 & 4.7\\
\hline \multicolumn{4}{c}{} \\[-1.5ex] \cline{1-6}
$\vartheta$ & 120 & 140 & 160 & 180 & 200 & \multicolumn{1}{c}{} \\
\cline{1-6}
$\sigma(\gamma)$ & 3.8 & 3.8 & 4.9 & 6.7 & 5.2 & \multicolumn{1}{c}{} \\
\cline{1-6} \multicolumn{4}{c}{} \\[-1.5ex] \cline{1-5}
$d$ & 0.1 & 0.2 & 0.3 & 0.4 & \multicolumn{2}{c}{} \\
\cline{1-5}
$\sigma(\gamma)$ & 1.8 & 2.7 & 4.9 & 9.0 & \multicolumn{2}{c}{} \\
\cline{1-5} \multicolumn{4}{c}{} \\[-1.5ex] \cline{1-5}
\phis & 0 & $-0.04$ & $-0.1$ & $-0.2$ & \multicolumn{2}{c}{} \\
\cline{1-5}
$\sigma(\gamma)$ & 4.9 & 4.9 & 4.9 & 5.4 & \multicolumn{2}{c}{} \\
\cline{1-5}
\end{tabular}
\caption{\captionFont Statistical uncertainty on $\gamma$ for one year of data. 
Unless otherwise specified, $\dms=20\invps$,
$\DGsGs =0.1$, $\gamma =65^\circ$, $\vartheta = 160^\circ$, $d =0.3$, 
$\phis = -0.04$.
\dms\ values are given in \invps, \phis\ values in radians, 
while $\gamma$, $\vartheta$ and $\sigma(\gamma)$ are given in degrees.
\label{tab:b2hh_gamma_sensitivity}}
\end{center}
\end{table}
The expected event yields and tagging performance for \prt{B \to
\pi\pi} and \prt{B_s \to KK} are given in table \ref{tab:BpipiYield}.
LHCb relies heavily on its \prt{K/\pi} separation capabilities to
achieve the required sample purity, as otherwise the different
hadronic two body decay modes of B hadrons are virtually
indistinguishable.

The statistical precision on $\gamma$ that can be achieved with
reconstruction and tagging performance depends on various parameters,
especially the ``penguin over tree ratio'', $d$, and the rapidity of
\prt{B_s} oscillations, given by the mass difference \dms.  For a
typical parameter set the precision on $\gamma$ is $\sim
\degrees{5}$. Statistical uncertainties on $\gamma$ for various sets
of parameters are given in table \ref{tab:b2hh_gamma_sensitivity}.
\subsection{\prt{B_s \to D_s K}}

An alternative way to tackle the problem of penguin contributions is
to look at decays that don't have any, like \prt{B_s \to D_s K}, or
\prt{B_d \to D^{(*)}\pi} \cite{DunietzBDK}. These decays are expected
to be rather insensitive to \np\ contributions, and therefore measure
a ``Standard Model $\gamma$'', providing a benchmark that other
decays, that are more sensitive to \np, can be compared against. Since
the final state is not a CP eigenstate, two CP-conjugate asymmetries
need to be measured,
\[ 
 A(\tau) = 
 \frac{     \Gamma \left(\prt{B_s \to D_s^- K^+}        \right) 
          - \Gamma \left(\prt{\bar{B}_s \to D_s^- K^+}  \right)
      }{    \Gamma \left(\prt{B_s \to D_s^- K^+}        \right) 
          + \Gamma \left(\prt{\bar{B}_s \to D_s^- K^+}  \right)
      }
\;,\;\;
 \bar{A}(\tau) = 
 \frac{     \Gamma \left(\prt{\bar{B}_s \to D_s^+ K^-}        \right) 
          - \Gamma \left(\prt{B_s \to D_s^+ K^-}  \right)
      }{    \Gamma \left(\prt{\bar{B}_s \to D_s^+ K^-}        \right) 
          + \Gamma \left(\prt{B_s \to D_s^+ K^-}  \right)
      }
\]
 The CP violating effect is in the difference between those
 asymmetries. This measurement is sensitive to $2\delta\gamma +
 \gamma$, and a possible strong phase difference
 $\Delta_{\mathrm{T1/T2}}$. Further details are given in
 \cite{DunietzBDK}.
\begin{table}[t]
\centering
\begin{tabular}{|@{}c@{}|c|c|c|c|c|c|}
\cline{1-5}
\dms  & 15 & 20 & 25 & 30 & \multicolumn{2}{c}{} \\
\cline{1-5}
$\sigma(\djdj)$ &  12.1 & 14.2 & 16.2 & 18.3 & \multicolumn{2}{c}{} \\
\cline{1-5} \multicolumn{3}{c}{} \\[-1.5ex]  \cline{1-4}
\DGsGs  & 0 & 0.1 & 0.2 & \multicolumn{3}{c}{} \\
\cline{1-4}
$\sigma(\djdj)$ & 14.7 & 14.2 & 12.9 & \multicolumn{3}{c}{} \\
\cline{1-4} \multicolumn{4}{c}{} \\[-1.5ex] \hline
\djdj  & 55 & 65 & 75 & 85 & 95 & 105 \\
\hline
$\sigma(\djdj)$ & 14.5 & 14.2 & 15.0 & 15.0 & 15.1 & 15.2 \\
\hline \multicolumn{4}{c}{} \\[-1.5ex] \cline{1-6}
$\Delta_{\mathrm{T1/T2}}$ & $-20$ & $-10$ & 0 & $+10$ & $+20$ &  \multicolumn{1}{c}{} \\
\cline{1-6}
$\sigma(\djdj)$ & 13.9 & 14.1 & 14.2 & 14.5 & 14.6 &  \multicolumn{1}{c}{} \\
\cline{1-6}
\end{tabular}
\caption{\captionFont Expected statistical uncertainty on $2\delta\gamma + \gamma$
for one year of data.  Unless otherwise specified, $\dms=20 \invps$,
$\DGsGs =0.1$, $2\delta\gamma + \gamma =\degrees{65}$ 
 and $ \Delta_{\mathrm{T1/T2}}=\degrees{0}$.  All values are
 given in degrees, except \dms\ in \invps.
\label{tab:BsDsKGamma}
}
\end{table}
 The particle ID capabilities of LHCb are crucial for the
 reconstruction of this decay, that would otherwise be swamped by
 background from \prt{B_s \to D_s^- \pi^+}, which has a $\sim 10$
 times higher branching ratio.  LHCb expects to reconstruct
 \un{5.4}{k} \prt{B_s \to D^- K^+} events per year with a
 background-to-signal of better than $0.5$.  This translates into a
 sensitivity on $\gamma$ of typically \degrees{\sim 15}, depending on
 other parameters, especially $\Delta m_s$. Results for different
 parameters sets are given in Table \ref{tab:BsDsKGamma}.

\subsection{$\gamma$ with \prt{B_d^0\to \bar{D}^0 K^{*0}}
                       \prt{B_d^0\to D^0_{\mathrm{CP}} K^{*0}}}
 The decay \prt{B_d^0\to \bar{D}^0 K^{*0}} offers the possibility of
 measuring $\gamma$, using untagged, time-integrated samples
 \cite{BDKstar}. This method is sensitive to New Physics in \prt{D^0}
 oscillations. The following 6 parameters are measured:
\\
\begin{tabular}{ccc}
\parbox{0.5\textwidth}{
\begin{eqnarray*}
\Gamma_+            &=& \Gamma\left(\prt{B^0 \to D^0(\pi^+ K^-)  K^{*0}}\right) \\
\Gamma_-            &=& \Gamma\left(\prt{B^0 \to \bar{D}^0(K^+ \pi^-)  K^{*0}}\right) \\
\Gamma_{\mathrm{CP}}&=& \Gamma\left(\prt{B^0 \to
                          D^0_{\mathrm{CP}}(K^+ K^-) K^{*0}}\right)
\end{eqnarray*}
}
&
\parbox{0.5\textwidth}{
\begin{eqnarray*}
\bar{\Gamma}_+            &=& \Gamma\left(\prt{\bar{B}^0 \to {D}^0(\pi^+ K^-)  \bar{K}^{*0}}\right) \\
\bar{\Gamma}_-            &=& \Gamma\left(\prt{\bar{B}^0 \to \bar{D}^0(K^+ \pi^-)  \bar{K}^{*0}}\right) \\
\bar{\Gamma}_{\mathrm{CP}}&=& \Gamma\left(\prt{B^0 \to
                          D^0_{\mathrm{CP}}(K^+ K^-) \bar{K}^{*0}}\right)
\end{eqnarray*}
}
\end{tabular}

They are related as follows:
\begin{equation}
{\Gamma_+ = \bar{\Gamma}_- \equiv g_1}, 
        \;\;\;{\Gamma_- = \bar{\Gamma}_+\equiv g_2}
\end{equation}
and
\begin{equation}
 \Gamma_{\mathrm{CP}}       = \frac{{g_1}+{g_2}}{2} + \sqrt{{g_1} {g_2}}
       \cos\left(\Delta + \gam\right)\;,\;\;
 \bar{\Gamma}_{\mathrm{CP}} = \frac{{g_1}+{g_2}}{2} + \sqrt{{g_1} {g_2}}
       \cos\left(\Delta - \gam\right)
\end{equation}
Where $\Delta$ is a possible strong phase difference.

\begin{table}
\begin{center}
\begin{tabular}{|l|@{\,\,}r@{\,\,}|@{\,\,}r@{\,\,}|@{\,\,}r@{\,\,}|@{\,\,}r@{\,\,}|@{\,\,}r@{\,\,}|@{\,\,}r@{\,\,}|}
\hline
$ \gamma$ & $55^\circ$ & $65^\circ$ & $75^\circ$ & $85^\circ$ & $95^\circ$ &
$105^\circ$ \\ 
\hline
$\sigma(\gamma)$ &  $9.0^\circ$ & $8.2^\circ$ & $7.6^\circ$ & $7.1^\circ$ & 
$7.0^\circ$ & $7.0^\circ$ \\
\hline
\end{tabular}
\caption{\captionFont Expected statistical precision on $\gamma$ for
different values of $\gamma$ after one year of data taking.
The value of $\Delta$ is set to $0$.
%
\label{tab:tn-tab-results}}
\end{center}
\end{table} 

 LHCb expects within $1$ year of data taking to reconstruct
 \un{3.6}{k} events to measure $\Gamma_-, \bar{\Gamma}_-$,
 \un{0.49}{k} events to measure $\Gamma_+, \bar{\Gamma}_+$, and
 \un{0.31}{k} events to measure $\Gamma_{\mathrm{CP}},
 \bar{\Gamma}_{\mathrm{CP}}$. Mainly because no tagging is required,
 the statistical weight of each reconstructed decay is much higher
 than in measurements using time-dependent decay rate
 asymmetries. Therefore, despite the comparably small data sample, a
 very competitive precision on $\gamma$ of $\sigma(\gamma) =
 \degrees{8.2}$ (for $\gamma=\degrees{65}, \Delta=0$) after one year
 can be achieved. Results for different values of $\gamma$ are given
 in table \ref{tab:tn-tab-results}.

\section{Summary}
\begin{table}
\begin{center}
\begin{tabular}{cc}
\\
\parbox{0.5\textwidth}{
\begin{tabular}{ l *{2}{c}}
Channel & Standard & New \\
        & Model    & Physics \\
\hline\hline
\prt{B_d^0\to D^{(*)\pm}\pi^{\mp}} & 
                                        \tick &      \\
\prt{B_s^0\to D_s^{\pm} K^{\mp}} & 
                                        \tick &      \\
\hline
\(
\parbox{2.6cm}{
\prt{B_d^0\to \pi^+\pi^-}\\
\prt{B_s^0\to K^+ K^-   }}\;\Bigg\}
\)& 
                                                  &\tick \\
\hline
\(
\parbox{2.6cm}{
\prt{B_d^0\to \bar{D}^0 K^{*0}}
\prt{B_d^0\to D^0_{\mathrm{CP}} K^{*0}}}\;\Bigg\}\)  &     &\tick \\
\end{tabular}\vspace{1ex}\\
}
&
\fbox{\parbox{0.4\textwidth}{ Typical precision in each channel
$\degrees{\sim 5}-\degrees{15}$ after 1 year.  More channels are under
investigation, e.g.: \(
\parbox{2.6cm}{
 \prt{B_d^0\to D_d^+ D_d^-}
 \prt{B_s^0\to D_s^+ D_s^-}}\;\Bigg\}\)\\
 which is highly sensitive to New Physics, since
 $\gam$ enters via penguins only \cite{FleischerDD}.
 (\prt{B_s^0\to D_s^+ D_s^-} is
 also sensitive to \DGsGs, and $\delta\gamma$).
}}
\end{tabular}
\caption{\captionFont Some $\gamma$ sensitive channels accessible at LHCb. It is
  indicated if the channels are expected to be sensitive to New
  Physics, or to the Standard Model $\gamma$.\label{tab:gammaSummary}}
\end{center}
\end{table}
 The recently re-optimised LHCb detector \cite{reotdr} is on track for
 data taking in 2007. The detector is designed to make best use of the
 vast number of B hadrons of all flavours, that are expected at the
 LHC.  LHCb has a comprehensive programme of high-precision B physics,
 including competitive measurements of the CKM phases $\bet, \gamma$,
 and mass and lifetime difference in the \prt{B_s} system within the
 first year of data taking. The physics programme also includes higher
 order effects (e.g. $\delta\gamma$), rare B decays, and many more.

 In this report we focused on one of the most exciting prospects at
 LHCb, the possibility to perform precision measurements of the angle
 \gam\ in many different decay channels in both the \prt{B_d} and
 \prt{B_s} system.  Some of the measurements will be more and some
 less sensitive to New Physics contributions. A selection of such
 channels are listed in Table~\ref{tab:gammaSummary}. The typical
 resolution in $\gamma$ is $\degrees{5}-\degrees{15}$ for each
 channel, within a single year of data taking.  This will thoroughly
 over constrain the Standard Model description of CP violation,
 providing important Standard Model measurements with a high
 sensitivity to New Physics.

%



\newcommand{\articleTitle}[1]{}

%

\end{document}